\documentclass[prd,superscriptaddress,preprintnumbers,amsmath,amssymb,nofootinbib]{revtex4}

\usepackage{amsmath}
\usepackage{amssymb}
\usepackage{slashed}
\usepackage{graphicx}
\usepackage{graphics}
\usepackage{epsfig}
\usepackage{color}
\usepackage{soul}
\usepackage{hyperref}
\usepackage[section]{placeins}
\usepackage{mathrsfs}
\usepackage{dsfont}

\newcommand{\be}{\begin{equation}}
\newcommand{\ee}{\end{equation}}
\newcommand{\bea}{\begin{eqnarray}}
\newcommand{\eea}{\end{eqnarrray}}

\newcommand{\Eqref}[1]{Eq.~\eqref{#1}}

\begin{document}
\title{Spectral dimension in causal set quantum gravity}

\author{Astrid Eichhorn}
\affiliation{\mbox{\it Perimeter Institute for Theoretical Physics, 31 Caroline Street N, Waterloo, N2L 2Y5, Ontario, Canada}
\mbox{\it E-mail: {aeichhorn@perimeterinstitute.ca}}}

\author{Sebastian Mizera}
\affiliation{\mbox{\it Girton College, University of Cambridge, Huntingdon Road, Cambridge CB3 0JG, United Kingdom}}

\begin{abstract} 
We evaluate the spectral dimension in causal set quantum gravity by simulating random walks on causal sets. In contrast to other approaches to quantum gravity, we find an increasing spectral dimension at small scales. This observation can be connected to the non-locality of causal set theory that is deeply rooted in its fundamentally Lorentzian nature.
Based on its large-scale behaviour, we conjecture that the spectral dimension can serve as a tool to distinguish causal sets that approximate manifolds from those that do not. As a new tool to probe quantum spacetime in different quantum gravity approaches, we introduce a novel dimensional estimator, the causal spectral dimension, based on the meeting probability of two random walkers, which respect the causal structure of the quantum spacetime. We discuss a causal-set example, where the spectral dimension and the causal spectral dimension differ, due to the existence of a preferred foliation.
\end{abstract}

\maketitle

\section{Introduction}
In quantum gravity, classical spacetime is replaced by a  quantum superposition of spacetimes, which can exhibit nontrivial properties. Here, much can be learned from the consideration of a fictitious, backreaction-free particle, probing a quantum spacetime by a random walk.
A first probe of quantum gravity effects is given by the spectral dimension, which is defined as follows: From the return probability  $P_r(\sigma)$ of the random walker to his starting point after diffusion time $\sigma$, averaged over the complete spacetime, one defines a generalized spectral dimension
\be
d_s(\sigma) =-2  \frac{\partial \ln P_r(\sigma)}{\partial \ln \sigma}.\label{d_s}
\ee
This is a measure of dimensionality that agrees with the topological or Hausdorff dimension for a flat smooth spacetime for $\sigma \rightarrow 0$:
For Brownian motion on a $d$-dimensional Riemannian flat background the probability density satisfies the diffusion equation $\left(\partial_\sigma - \nabla_x^2\right)P(x,x',\sigma)=0$. The solution for the initial condition $P(x,x',0)= \delta^d(x-x')$ is then given by $P(x,x',\sigma) = (4 \pi \sigma)^{-d/2}e^{- \vert x - x' \vert^2/ (4 \sigma)}$. Accordingly $P_r(\sigma)=\frac{1}{V} \int d^d x  P(x,x,\sigma) =\left( 4 \pi \sigma\right)^{-d/2} $ holds, and thus $d_s(\sigma \rightarrow 0)=d$. In contrast, the topological and the spectral dimension can differ in quantum gravity settings. The microscopic structure of spacetime can change the small-$\sigma$ behaviour of $d_s$, whereas the infrared, semiclassical limit, where $d_s = d$ must hold, is reached for large $\sigma$.
This has been observed in Causal Dynamical Triangulations (CDT) \cite{Ambjorn:2005db,Ambjorn:2005qt,Gorlich:2011ga,Benedetti:2009ge,Anderson:2011bj} and multigraph models derived from CDTs \cite{Giasemidis:2012rf,Giasemidis:2012qk}, asymptotically safe gravity \cite{Lauscher:2005qz,Lauscher:2005xz,Reuter:2011ah,Rechenberger:2012pm,Calcagni:2013vsa} and more recently in Ho\v{r}ava gravity \cite{Horava:2009if,Sotiriou:2011mu,Calcagni:2013vsa}, where a dimensional reduction from $d_s=4$ to $d_s=2$ is observed in $d=4$. Indications for the same phenomenon also exist in Loop Quantum Gravity \cite{Modesto:2008jz,Calcagni:2013vsa}, in settings with space-time noncommutativity \cite{Benedetti:2008gu} and in a  model of nonlocal gravity \cite{Modesto:2011kw}. The strong-coupling limit of the Wheeler-DeWitt equation also shows a similar behaviour \cite{Carlip:2009kf}, and possible connections to Belinsky-Khalatnikov-Lifshitz-type behaviour and quantum-gravity induced focusing of light cones have been discussed in \cite{Carlip:2011tt}. An effective description in terms of fractional geometry has been attempted in \cite{Calcagni:2012rm}.

To extract the spectral dimension from a quantum gravity model, two different routes have been followed: One can either consider the flat-space diffusion equation, and modify the Laplacian accordingly to include quantum gravity effects \cite{Lauscher:2005qz}, by substituting $\nabla^2 \rightarrow \langle \nabla^2 \rangle$. A second route is open in quantum gravity models where spacetime can be represented as a graph,
consisting of vertices and edges, such as in the case of CDTs, causal sets, or spin foams. There one can study the diffusion process itself on the graph, and infer a spectral dimension from the return probability, see, e.g. \cite{Durhuus:2009zz} and \cite{Ambjorn:2005db,Ambjorn:2005qt,Gorlich:2011ga,Benedetti:2009ge,Anderson:2011bj}. This procedure can also be applied to a Lorentzian quantum gravity model, where it is nontrivial to set up a well-defined diffusion equation \cite{Debbasch}.

Here, we will present a first exploration of  the spectral dimension in causal set quantum gravity.  The causal set approach is based on Lorentz invariance and discreteness of quantum spacetime. A quantum spacetimes in this setting corresponds to a set of elements (spacetime points), with a relation that encodes the causal order represented by links between the elements. This partial order can be represented by  a graph.
We will  consider the random walk on that graph to extract the spectral dimension. 
The combination of Lorentz invariance and discreteness in causal sets entails a particular form of nonlocality, which will have profound consequences for the small $\sigma$ behaviour of the spectral dimension, as we will see.  In contrast to other quantum gravity approaches, we will observe an increase of the spectral dimension at small scales.

This paper is structured as follows. We introduce the basic kinematic structure of causal set quantum gravity in sec.~\ref{causet}. We then discuss analytic expressions and simulations for a random walk on different causal sets, including $d=2,3,4$ dimensional Minkowski spacetime, $1+1$ dimensional spatially compact spacetime, $2d$  de Sitter spacetime, Kleitman-Rothschild orders and transitive percolation models in sec.~\ref{ds}. We then introduce a measure of dimensionality from the meeting probability of two random walkers respecting causality in sec.~\ref{meetdim}, and conclude in sec.~\ref{conclusion}.

\section{Causal set quantum gravity}\label{causet}
Causal set quantum gravity is a fundamentally Lorentzian approach to quantum gravity, that is kinematically discrete. 
The theory's basic postulate is that the fundamental nature of spacetime is a locally finite causal order \cite{Bombelli:1987aa}.  
The dynamics is  defined as a path-integral over causal sets, that, in common with some other approaches to quantum gravity see, e.g. \cite{Rivasseau:2012yp,Di Francesco:1993nw},  includes quantum fluctuations of topology as well as fluctuations of geometry. A proposal for the microscopic quantum dynamics in the form of a family of microscopic actions that approximate the Einstein-Hilbert action in the discrete setting has been made in \cite{Benincasa:2010ac,Benincasa:2010as, Dowker:2013vba},  allowing to evaluate the path-integral over causal sets, \cite{Surya:2011du}. For reviews on the causal set program, see, e.g. \cite{Henson:2010aq,Dowker:2005tz,Sorkin:2003bx,Dowker:aza,Surya:2011yh}.

More precisely, a causal set (causet for short) $\mathcal{C}$ is a set of elements (which are spacetime points), with a relation $\prec$, called ``precedes'', which encodes the causal structure, thus containing  part of the information that is carried by the metric in the continuum. It satisfies the following axioms:
\begin{itemize}
\item transitivity: If $x \prec y$ and $y \prec z$, then $x \prec z$, $\forall x,y,z \in \mathcal{C}$.
\item non-circularity: If $x \prec y$ and $y \prec x$, then $x = y$, $\forall x,y \in \mathcal{C}$.
\item local finiteness: $\forall x, z \in \mathcal{C}$ the set $\{y \, \vert \, x \prec y \prec z, y \in \mathcal{C} \}$ is finite.
\end{itemize}
The first obviously holds for a causal relation, and the second forbids closed timelike curves.
The third axiom encodes the fundamental kinematical discreteness of causal sets:
 unlike in the continuum, where an infinite number of points lie between any two given spacetime points, there is only a finite number of these in a causal set. Points at spacelike separation are unrelated in a causal set, thus the causal set corresponds to a partial order.

As the causal structure does not encode information on the conformal factor, the volume information contained in the metric is so far missing. Assigning a mean spacetime volume of $l_{\rm Pl}^d$ to each element of a causal set then recovers volume information from counting causal set elements, thus the causal set slogan: ``order+number = geometry'' \cite{RSorkin1, RSorkin2}. Here $l_{\rm Pl}=1$ is the fundamental discreteness scale, which a priori could be any scale, but most naturally lends itself to an identification with the Planck scale.

On a causal set, two distinct types of relations exist: We will call links those fundamental relations that are not implied by transitivity, i.e. those connecting ``nearest neighbors'' (where nearness refers to a notion of proper time). They are distinct from those relations implied by transitivity and constitute the transitively reduced causal set. This is represented by the Hasse diagram, which will form the central object of our investigations.

The causal structure of spacetime is  nonlocal in the sense that short distances in the Lorentzian setting can translate to large distances in a Wick-rotated setting. Therefore causal sets are intrinsically nonlocal: In a causal set that is approximated by Minkowski spacetime, any point will have an infinite number of nearest neighbors. In any given frame, most of these will lie at large spatial distance \cite{Moore:1988zz,Bombelli:1988qh}. This form of nonlocality will play a crucial role in our analysis. Note that spacetime curvature can tame the causal set nonlocality: For instance in a Friedman-Robertson-Walker universe a finite number of past links exist \cite{Bombelli:1988qh,BSchmitzer}.

In causal set quantum gravity background structure such as dimensionality or topology is given up, thus in the set of all causets with $N$ elements, most causal sets do not approximate manifolds. Then the recovery of a semiclassical limit from the causal set path integral is particularly challenging and requires a nonlocal dymanics \cite{Benincasa:2010ac}. In particular, it is necessary to develop causal set estimators of continuum properties, such as the topology or the dimensionality, see, e.g.  \cite{Myrheim:1978ce,Meyer,Bombelli:1987vp,Reid:2002sj, Glaser:2013pca} for estimators of the topological dimension.

\subsection{Sprinklings: Constructing causets approximated by Lorentzian manifolds}
If the fundamental structure of spacetime is a causal set, a smooth manifold emerges as a continuum approximation on semiclassical scales. As a crucial difference to, e.g. the CDT programme, causal sets are fundamentally discrete. Therefore it is not the continuum limit ($l_{\rm Pl}/l \rightarrow 0$), but the continuum approximation ($l_{\rm Pl}/ l\ll1$) that is of interest. Thus the calculation of physical quantities does not require the tuning to a second-order phase transition in the parameter space. In other words, the fundamental scale $l_{\rm Pl}$ carries a physical meaning in this case, and is not introduced as an unphysical regularisation. Observables may therefore depend on $l_{\rm Pl}$ or $N$, the total number of elements in a given causal set.  We expect that if causal sets yield a realistic description of gravity, the long-distance features will approach continuum properties of (semi)classical spacetimes.

In order to find a causal set which will be approximated by a given manifold, we will employ the method of sprinklings \cite{Bombelli:1987aa}, which is based on a Poisson process: Taking a given volume $V$ in $(d+1)$ dimensional spacetime, the probability to find $n$ causal-set elements in it is given by
\be
\mathcal{P}(n, V)= \frac{1}{n!}(\rho V)^n e^{- \rho V},\label{Poisson}
\ee
where $\rho$ is the sprinkling density.
This generates a distribution that does not exhibit a preferred frame and is thus invariant under Lorentz symmetry \cite{Bombelli:2006nm}, as only a random distribution -- in contrast to a regular lattice -- can be.

A crucial property of a causal set approximating $d+1$ dimensional Minkowski spacetime is that each point has an infinite number of nearest neighbours, i.e., points it shares a link with. 
This property crucially hinges on the combination of Lorentz invariance and discreteness: Discreteness implies that in any given region $R$ that is not spacelike to a given point $x$, there is a finite probability $p$ of finding $n$ elements that share a link with $x$. Lorentz invariance implies that if we boost this region $R$, we will obtain a region $R'$ that is disjoint with $R$, and  must have the same probability $p$ to find $n$ elements with links to $x$. In this way be obtain an infinite number of disjoint regions and accordingly infinitely many links to $x$ \cite{Moore:1988zz,Henson:2010aq}. Of these links, many will be to points at a large spatial distance in any given frame, thus making causets non-local.

To generate a sprinkling into a region of finite volume contained in Minkowski spacetime, we first draw the number of points from the Poisson distribution \Eqref{Poisson}, and then independently choose the coordinates for each of them according to a uniform distribution. The causal relations are those induced by the underlying Minkowski spacetime, cf. fig.~\ref{sprinkle_m2}. In a first step, all causal relations are included for any time- or lightlike pair (the second has, strictly speaking, measure zero in a Poisson sprinkling), before the causet is transitively reduced.

Similarly, we can construct a sprinkling into 1+1 dimensional curved spacetime by transforming the metric to the conformally flat form $g_{\mu \nu} = \Omega^2(x) \eta_{\mu \nu}$, with $\eta_{\mu \nu}= \mathrm{diag}(-1,+1)$ 
and $\Omega^2(x)$ being the position-dependent conformal factor. Then, the sprinkling density needs to be rescaled by the conformal factor, which can be achieved by either constructing an appropriate mapping \cite{BSchmitzer} or rejecting excessive points \cite{Reid:2002sj}. Since the metric is conformally flat the causal links can be deduced in the same way as for Minkowski spacetime.

\section{Spectral dimension for causal sets}\label{ds}
We will now consider given classes of causal sets and study diffusion processes on these, without taking into account causal set dynamics. The spectral dimension realised in a universe described by causal set quantum gravity would be derived from the weighted return probability $\int_{\mathcal{C}} e^{i S[C]} P_r(\sigma; C)$, where $C$ denotes the causal sets that are summed over in the path-integral. Performing or approximating the path-integral is a highly challenging task \cite{Surya:2011du} which we will not attempt here.
If the dynamics implies that the expectation value over causal sets from the path integral will be a de Sitter sprinkling and the fluctuations are small, our results in sec.~\ref{deSitter} could be considered as the spectral dimension emerging from the path-integral for causal sets. Even without this dynamical information, the spectral dimension already shows very interesting behaviour.

Here, we will take causal sets as graphs, which can be probed by a random walker very much in the spirit of \cite{Durhuus:2009zz}. A random walker follows a walk, consisting of a set of causal set elements connected by links. This is in analogy to the investigations in \cite{Ambjorn:2005db,Gorlich:2011ga,Benedetti:2009ge,Anderson:2011bj}, where the spectral dimension of CDTs was determined by a random walk jumping from one simplex to the neighbouring simplices in a triangulation. A crucial difference to that setting is that steps which are purely spacelike
are not possible on a causal set, where every step is timelike (lightlike steps have measure zero in a Poisson sprinkling). 

The probability $p(x,x')$ to move from a point $x$ to a point $x'$ in a causet of finite size is then given by
\be
p(x,x') =\begin{cases}
1/n(x) &\quad \text{if } x \text{ and } x' \text{ are linked}, \\
   0   &\quad \text{otherwise}, 
  \end{cases}
\ee
where $n(x)$ is the number of links starting or ending at $x$, i.e. the degree of $x$. At a given point $x$, the next point of the random walk is chosen uniformly at random from the neighbours, i.e. those points sharing a link with $x$. This random walk 
does not respect the causal ordering, and diffusion can occur into the future as well as into the past. This underlines that the diffusing particle in this setting is a fictitious probe particle, and this random walk is unrelated to physical propagation of particles on a quantum spacetime\footnote{Interestingly, it has been suggested that physical particle propagation on causal sets can also be related to a particular diffusion equation \cite{Dowker:2003hb,Philpott:2008vd}.}. Note that no diffusion can occur between points at spacelike separation, as there cannot be a link between them. In this way, we get a positive definite diffusion probability even though we choose it to be proportional to the proper distance (a link on average corresponds to a proper distance of $1$). This circumvents one of the arguments why a diffusion equation on Lorentzian backgrounds is challenging to set up in the continuum, as a choice of the probability proportional to the proper time would correspond to ``negative probability'' for spacelike separations \cite{Debbasch}. Restricting the diffusion to take place within the lightcone resolves this problem.

The probability for a walk from $x$ to $x'$ in $n$ steps is then obtained as
\be
P(x,x',n)= \begin{cases}
\sum_{p(x,x')}\Pi_{i=1}^{n}\frac{1}{n(x_i)} &\quad \mbox{ if a path $p(x,x')$ in $n$ steps exists from $x$ to $x'$, consisting of the points $x_i$},\\
0 &\quad \mbox{ otherwise}.
\end{cases}
\ee
As different paths between $x$ and $x'$ constitute mutually exclusive cases, we have to sum over all possible paths. Note that for all paths from $x$ that have $n$ steps, $\sum_y\sum_{p(x,y)} \Pi_{i=1}^n \frac{1}{n(x_i)}=1$.

To evaluate the spectral dimension from $P(x,x,n)$ we must average over the starting point of the random walk. A further averaging procedure over all sprinklings into a given spacetime yields the spectral dimension associated to a class of causal sets that approximate a given spacetime.

On a discrete, lattice-like structure, a random walk will generically show the following behaviour for small diffusion times: In the first step, the return probability is zero, and becomes non-zero in the second step. Depending on the structure of the lattice, the return probability will be zero for the third step again, e.g. in the case of a two-dimensional square lattice. This oscillating behaviour of the return probability is a clear signature of a discrete setting. It implies that if the random walk is split into even and odd timesteps, and a spectral dimension is derived from the two cases, its value will differ.
For larger diffusion times, this behaviour is ``washed out'' in many settings, and the return probability becomes an approximately smooth function of the diffusion time.\footnote{This effect of the discreteness can in principle be removed by introducing a parameter that encodes the probability to remain at the same point \cite{Benedetti:2009ge}.} 

To explicitly evaluate the spectral dimension, two routes are open to us, which should give the same results: Firstly, we can evaluate the spectral dimension from simulations or analytical studies of random walks on the causet, i.e., on the graph that defines it, the Hasse diagram. Secondly, we could use the embedding information in the case where the causet is approximated by a smooth Lorentzian spacetime, and derive the spectral dimension in the continuum approximation. In that case, we obtain an expression that seems rather challenging to tackle in practice \footnote{Initially the probability is peaked at the starting point and given by $P(x, x_0,0)= \delta^{d+1}(x-x_0)$.
In the next step, the random walker can jump to a given point $x_1$ if there is a link to this point. This requires, that there is a causal set element at this point, and the causal volume between the two points $x_1$ and $x_0$ is empty of causet elements (otherwise there would not be a direct link). Using the Poisson distribution, we thus obtain
\be
P(x_1,x_0, 1) = A_0\, \rho\, e^{-\rho V(x_0,x_1)},
\ee
where $V(x_0,x_1)$ denotes the volume of the causal interval between $x_0$ and $x_1$ and $P(x,x_0,1)=0$ if there is no link. $A_0$ is a normalisation constant for $x_0$ to ensure that the integrated probability is 1. In the next step, jumping to point $x_2$, the probability density must involve an integral over all possible intermediate points $x_1 \in V_1 =(J^{+}_0 \cup J^{-}_0) \cap (J_2^+ \cup J_2^-)$, where $J^{\pm}_x$ denotes the future/past of $x$, so that
\be
P(x_2, x_0, 2)= \int_{V_1} d^{d+1} x_1 \sqrt{-g}\, A_1\, A_0\, \rho^2\, e^{- \rho V(x_0,x_1) - \rho V^{\ast}(x_1,x_2)}.
\ee
This makes the construction principle clear. The only further complication arises, as the volumes other than $V(x_0,x_1)$ have to exclude the regions that have been set to be devoid of any elements in previous steps of the walk. Hence, by $V^{\ast}(x_i, x_{i+1})$ we will denote the volume $V(x_i, x_{i+1})$ minus the intersections with the $V(x_j, x_{j+1})$, $j<i$.
The probability density after $n+1$ steps will accordingly read
\be
P(x_{n+1}, x_0, n+1)=\int_{V_n^\ast} d^{d+1} x_n \sqrt{-g} A_n \cdots \int_{V_1} d^{d+1} x_1 \sqrt{-g} A_1 A_0 \rho^{n+1} e^{-\rho V(x_0, x_1)- \ldots - \rho V^{\ast}(x_n, x_{n+1})}.
\ee
}. We thus focus on simulations of random walks on causets in the following.

\subsection{Spectral dimension of sprinklings into Minkowski spacetime}
\begin{figure}[!h]
\includegraphics[width=0.4\linewidth]{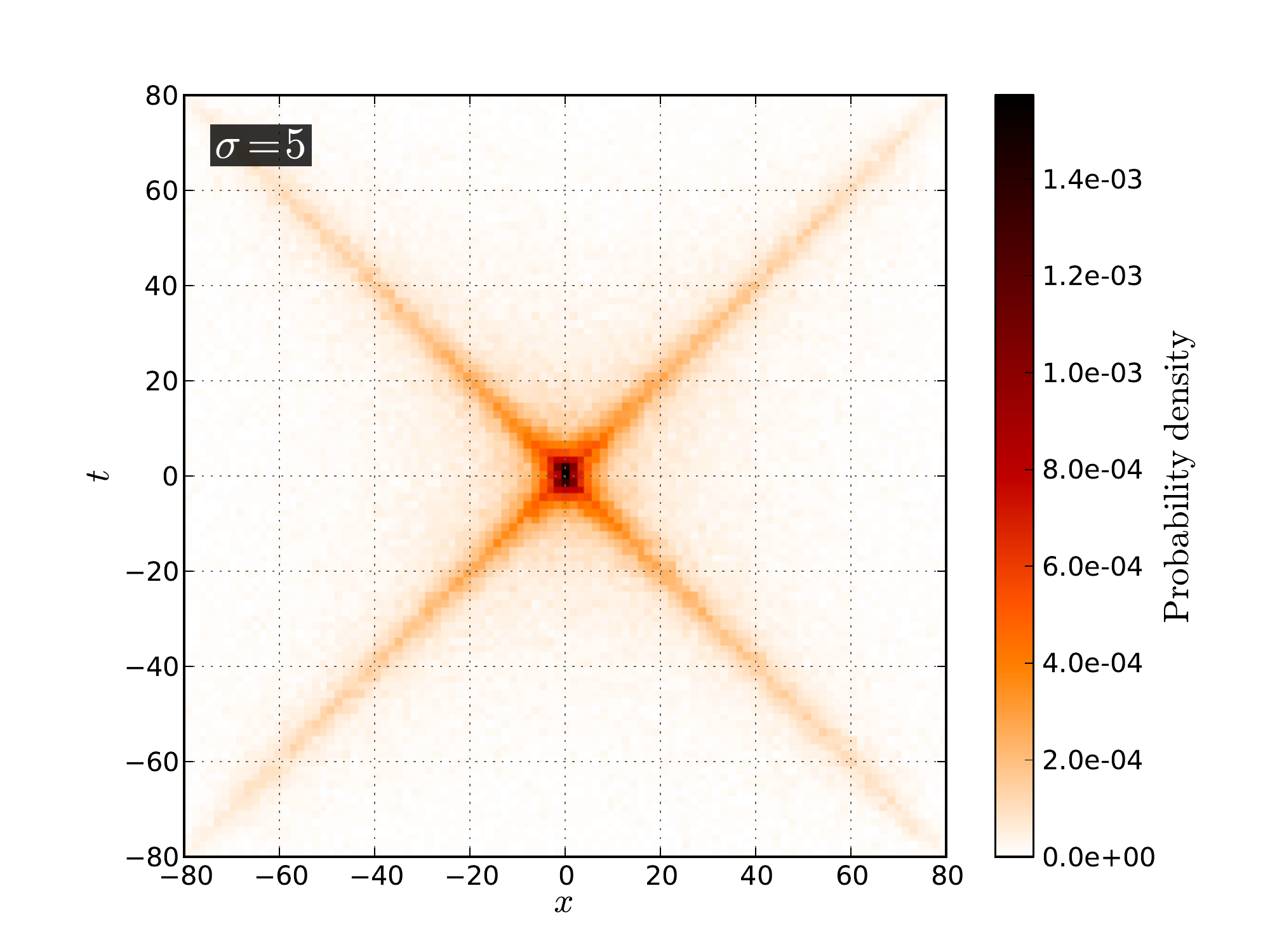}
\includegraphics[width=0.4\linewidth]{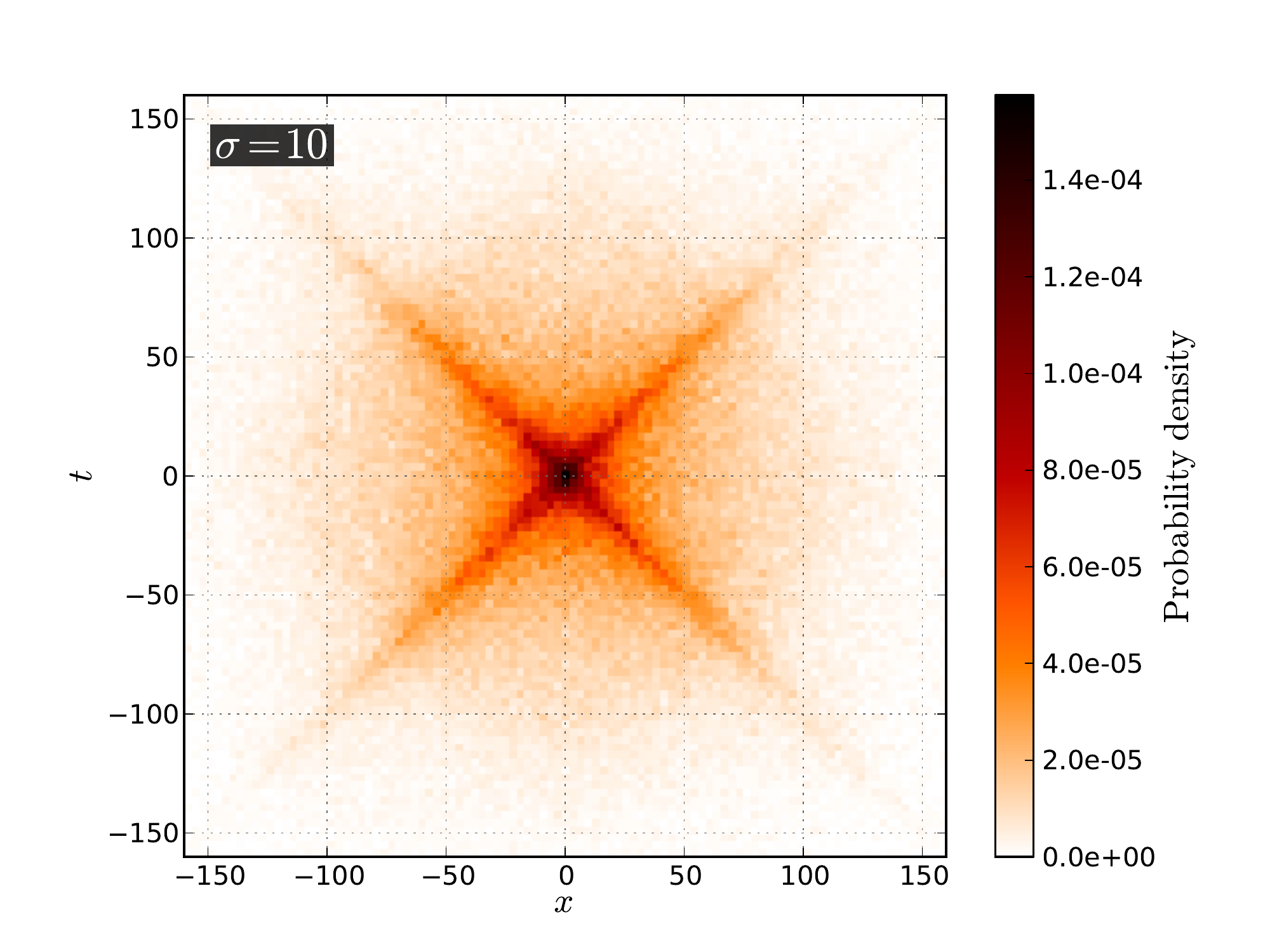}
\includegraphics[width=0.4\linewidth]{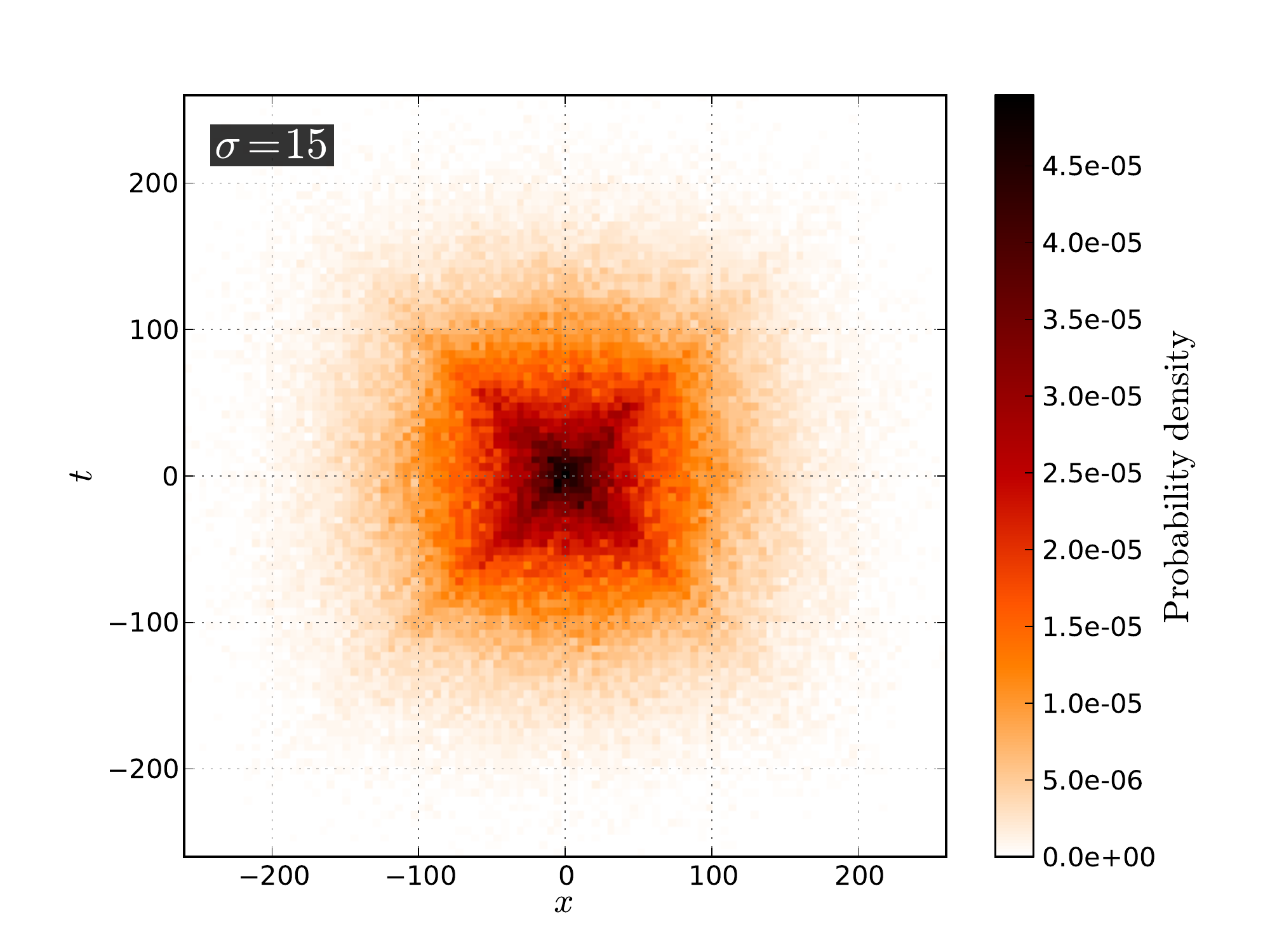}
\includegraphics[width=0.4\linewidth]{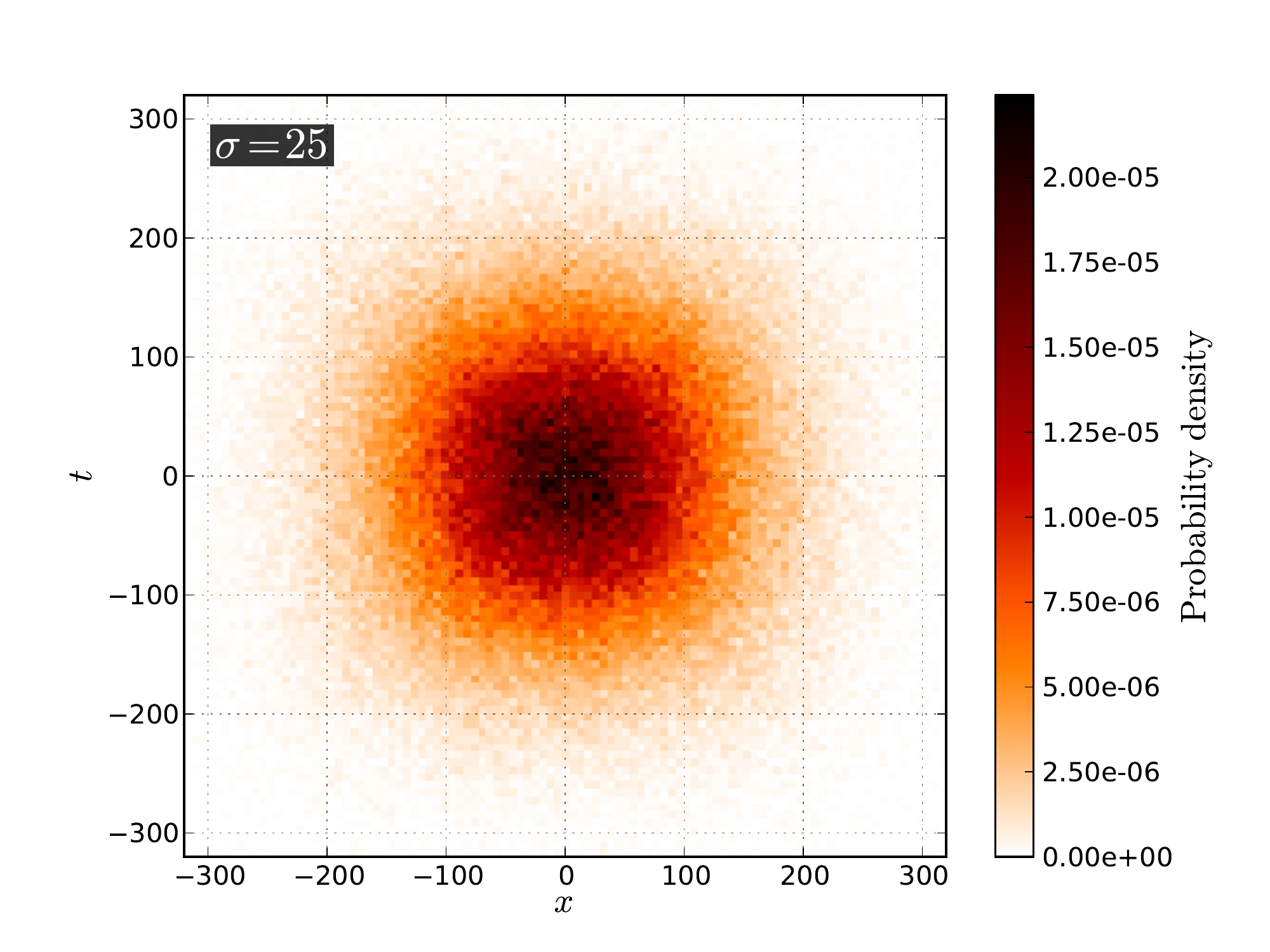}
\caption{\label{snap_m2}{Probability distribution function of a diffusion process on $1+1$ Minkowski sprinklings 
with cutoff length $L=100$ at diffusion times $\sigma = 5, 10, 15, 25$. The two stages of the process are visible: low-$\sigma$ behaviour reflects the non-local nature of causal sets and yields a high value of the spectral dimension, whereas the diffusion at longer times ($\sigma \gtrsim 20$) starts to resemble the solution to the heat equation on a smooth manifold, explaining the observed asymptotic value of $d_s = 2$. Note that the colours encoding the probability correspond to different values in the different snapshots. }}
\end{figure}

In $d+1$ dimensional Minkowski spacetime, the number of links from a given point in a causet of $N$ elements is \cite{Bombelli:1988qh}
\be
\# {\rm links} \sim  N^{(d-1)/2} \quad \mbox{($\log N$ for $d=1$)}.\label{linkno}
\ee
The number of nearest neighbours thus contains global and not just local information, as $N V_{\rm Pl} = V$, where $V$ is the total volume of the spacetime approximating the causet.
This property reflects the causal-set nonlocality, and is responsible for a drastic departure of the spectral dimension from other quantum gravity settings: Due to the nonlocality a random walker will move away from the origin very quickly, cf. fig.~\ref{snap_m2}, keeping close to the lightcone.

\begin{figure}[!here]
\includegraphics[width=0.35\linewidth]{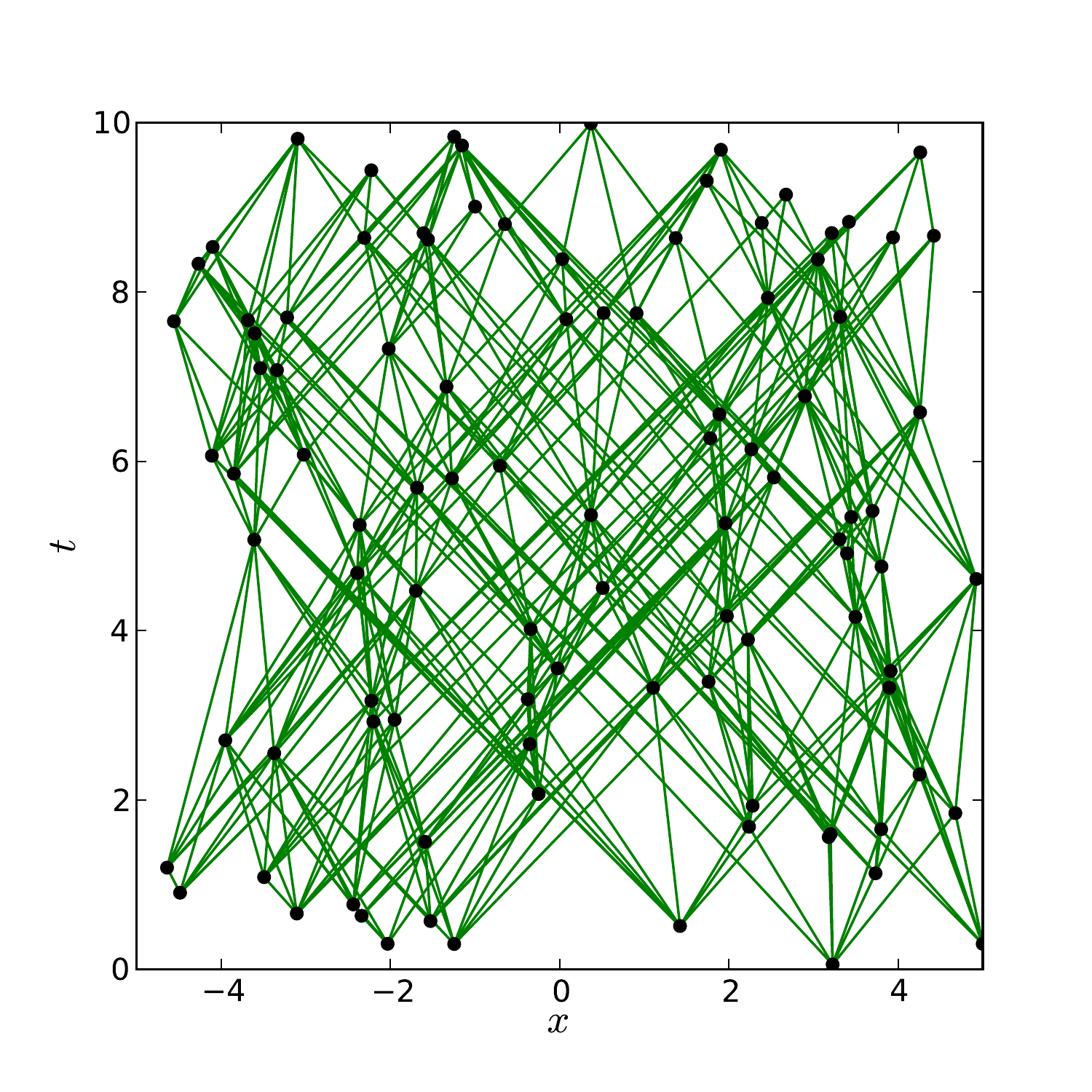}
\includegraphics[width=0.35\linewidth]{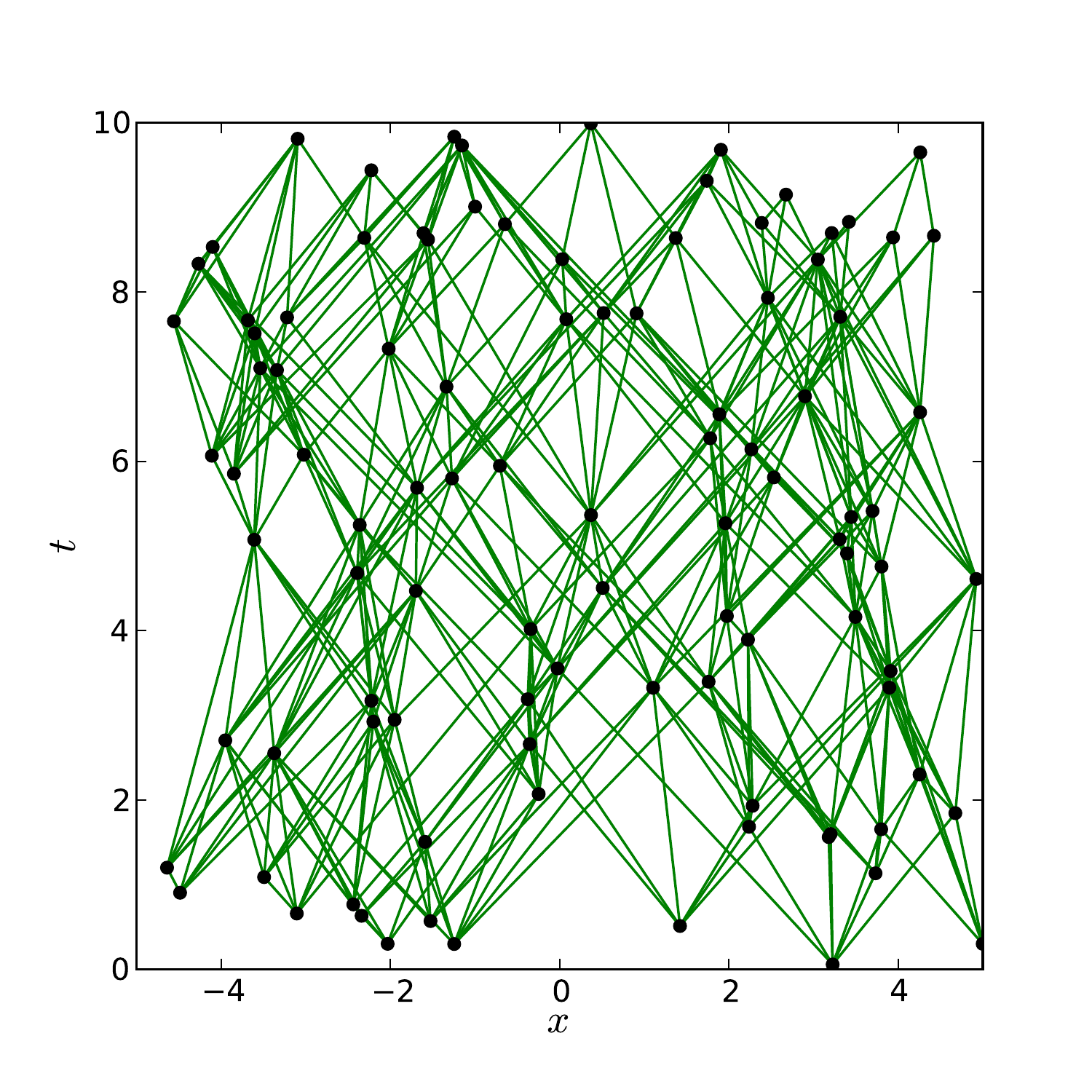}
\caption{\label{sprinkle_m2}{Sprinklings into $1+1$ dimensional Minkowski spacetime with unit density: without any cutoff (left), and with infrared cutoff $L=5$ (right). The causet elements and links are shown.}}
\end{figure}

More specifically, the nonlocal structure of causal sets implies that within a small number of steps
the random walker can get to a point where a typical return path consist of a large number of links, cf. fig. \ref{sprinkle_m2}. This is very different from the setting of, e.g. CDT, which is essentially local and where the number of neighbouring simplices is finite.
The consequence of the nonlocality is a strong change of the return probability in the first steps, yielding a large or even divergent spectral dimension. This follows, as the return probability is related to the inverse of the number of points at which the random walker can be at time $\sigma$. In a nonlocal setting, this number grows quickly, accordingly yielding a strong fall-off of $P_r(\sigma)$, resulting in a large $d_s$.
This behaviour is in marked contrast to other approaches to quantum gravity: There, the small-scale spectral dimension undergoes a dynamical reduction \cite{Ambjorn:2005db,Lauscher:2005qz, Modesto:2008jz, Horava:2009if, Calcagni:2013vsa}. In terms of the diffusion process this implies the property of subdiffusion, where the mean squared displacement grows slower with diffusion time  than in the case of Brownian motion on a flat background. Physically, one could imagine quantum fluctuations of spacetime ``hindering'' the progress of the diffusing particle.
Here, we report the first case of superdiffusion in a quantum gravity approach. It is rooted in the fundamentally Lorentzian nature of causal sets which, combined with discreteness, yields the strong nonlocality. This allows the particle to cover larger distances (in a Euclidean sense) than in other settings, thus showing superdiffusion.

In \cite{Modesto:2009qc} it has been postulated that the presence of a smallest length scale in a Riemannian setting should lead to a decrease of the spectral dimension at small scales. Here we provide an example for the Lorentzian setting, where a fundamental length scale exists, which serves as a Lorentz-invariant cutoff, as each element of a causal set corresponds to a Planckian volume on average. The existence of this fundamental length scale together with the requirement of Lorentz invariance and discreteness actually leads to an increase of the spectral dimension on small scales.

\begin{figure}[!h]
\begin{minipage}{0.5\linewidth}
\includegraphics[width=\linewidth]{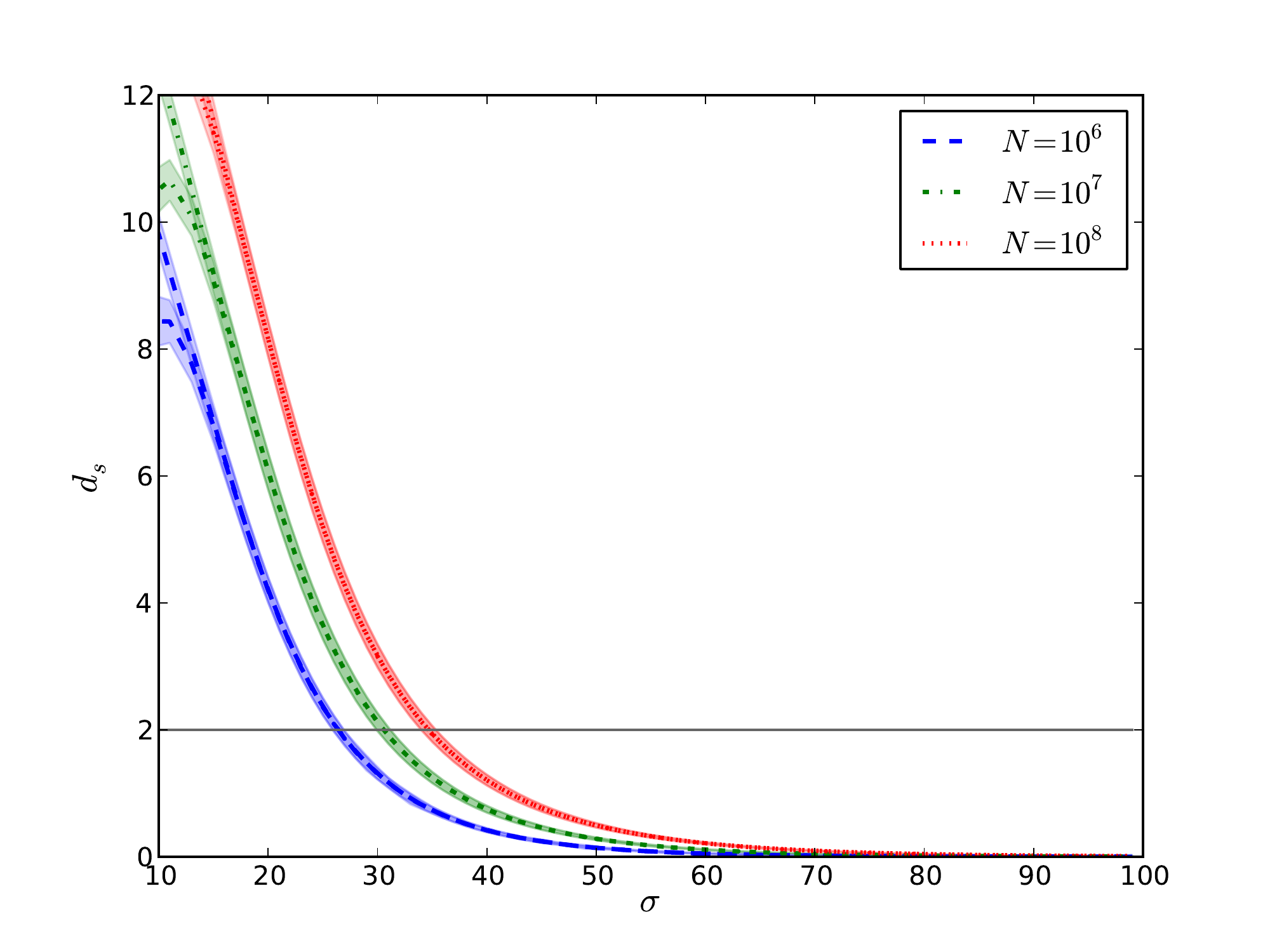}
\end{minipage}
\begin{minipage}{0.48\linewidth}
\caption{\label{m2_without_cutoff}Spectral dimension $d_s$ as a function of diffusion time $\sigma$ for sprinklings into $1+1$ dimensional Minkowski spacetime. Lines for three different causet sizes are shown, $N = 10^6, 10^7, 10^8$ with bands showing the $1\sigma$ statistical error. On average a random walker reaches a region close to the boundary within $\sigma \approx 30$ steps for the largest simulations. Accordingly the subsequent decrease of the spectral dimension is a boundary effect.  As discussed above, the small-$\sigma$ behaviour will show a difference between $d_s$ derived from only the even or only the odd steps of a random walk, as has been observed also in CDT \cite{Benedetti:2009ge}. The two curves converge quickly on sprinklings into Minkowski spacetime. Instead of performing single random walks we compute the exact probabilities at each timestep, thus obtaining the precise values of the return probability which allows to calculate its logarithmic derivative with less uncertainty. We found it enough to average over $\sim 100$ sprinklings and starting points to achieve convergent results. Very similar volumes and averaging numbers will be used in further simulations unless stated otherwise.}
\end{minipage}
\end{figure}

At larger scales in our simulations, the spectral dimension drops to zero quickly. This happens on finite causets, as the random walker quickly reaches the boundary, and is ``reflected''. The probability distribution then equilibrates, yielding a constant return probability and a vanishing spectral dimension. This drop off of the spectral dimension due to the boundary is also well-known from CDT  simulations \cite{Benedetti:2009ge}. In the case of causal set quantum gravity, the boundary effect is particularly severe: As explicit simulations of random walks can only be considered on finite sprinklings, there is always a finite probability to reach the boundary in a single step. Although this probability decreases with the number of points in the sprinkling, 
even for large sprinklings the spectral dimension will 
approach zero in a small number of diffusion steps, cf. fig.~\ref{m2_without_cutoff}.

The largest causets that we are able to treat due to limited computational resources have $N \approx 10^8$ elements, which corresponds to the volume of the spacetime in fundamental units. This is to be compared to a maximum volume of $2 \times 10^5$ used in the most recent CDT simulations \cite{Benedetti:2009ge}, which has proven to be large enough to avoid boundary effects. The finite, and volume independent, number of nearest neighbours in CDT in contrast to a number of nearest neighbours that grows with the volume in the case of causal sets is responsible for this major difference.

The same property implies that the small-scale value of the spectral dimension depends on $N$: Due to the nonlocality of a causet a random walker has access to the information on the total size of the causal set within the very first steps, due to \Eqref{linkno} This is crucially different from local settings, where the local movement of the random walker is independent of the total size of the quantum universe that he probes. In the causet case this dependence on $N$ is a physical one. This differs from settings where a discretisation is introduced as a regularisation, and is unphysical. Then the number of building blocks must diverge in the physical continuum limit. This limit is obtained by tuning to a second-order phase transition, where the correlation length diverges. Thus the finite size of the system will affect the results.
A finite-size scaling analysis is then in order, to obtain the continuum result from a finite simulation. In the case of causal sets, the discretisation is physical. $N$ is therefore a physical number, related to the number of spacetime points in our universe. Combined with the nonlocality of causal sets this implies that the dependence of $d_s$ on $N$ at small $\sigma$ is physical. Away from a second-order phase transition, finite-size effects in lattice-like systems do usually not play an important role, but in the causal set case, the nonlocality implies that the system size $N$ matters, and affects the small-scale value of $d_s$.

To analyze the sprinklings into Minkowski spacetime systematically, we will now introduce an infrared cutoff, see fig.~\ref{sprinkle_m2}. Its main purpose is to move the onset of the boundary effect to larger $\sigma$, thus revealing a non-trivial intermediate-$\sigma$-regime for $d_s$.

This cutoff is chosen in a particular frame, where we use the Wick-rotated Euclidean metric to discard links stretching beyond the infrared cutoff distance $L$. 
The effect of the cutoff is to reduce the causal set non-locality, and make it resemble a local lattice structure with a bound on the maximal number of links at a given point even for a sprinkling into Minkowski spacetime without boundaries, see fig.~\ref{sprinkle_m2}. The cutoff can thus be understood as a cutoff on the degree of the vertices, which is infinite on a Minkowski spacetime. Its additional effect is to reduce the boundary effect: Introducing a cutoff we can explicitly simulate random walks on the causal set where the random walker will not reach the boundary of the sprinkling within a given number of steps. This allows us to probe a regime of intermediate $\sigma$ without boundary effects. Without a cutoff, much larger causets would be necessary to access this regime, requiring significantly more computational resources.
Besides, using a cutoff can be understood as an approximation of the possible effect of spacetime curvature: As curvature affects the global light-cone structure, 
curvature can affect the number of links \cite{Bombelli:1988qh}. Other settings with a finite number of links include finite-volume spacetimes, such as in the case of compact extra dimensions.
Taking the limit $L \rightarrow \infty$ provides a controlled way to recover the full causal set and its non-locality. In that limit, the preferred frame selected by the cutoff is removed and Lorentz invariance is restored.

\begin{figure}[!here]
\includegraphics[width=0.48\linewidth]{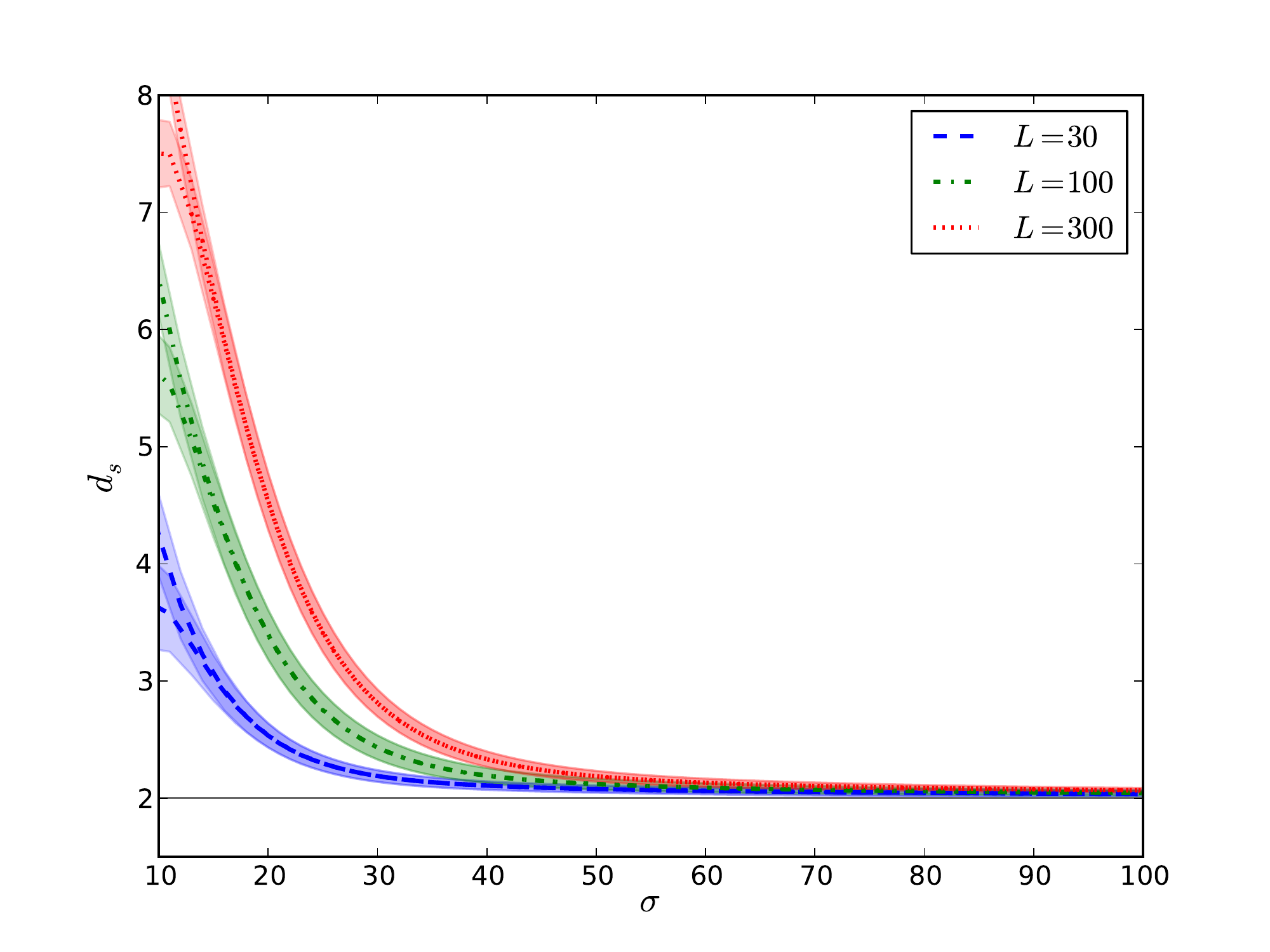} \includegraphics[width=0.48\linewidth]{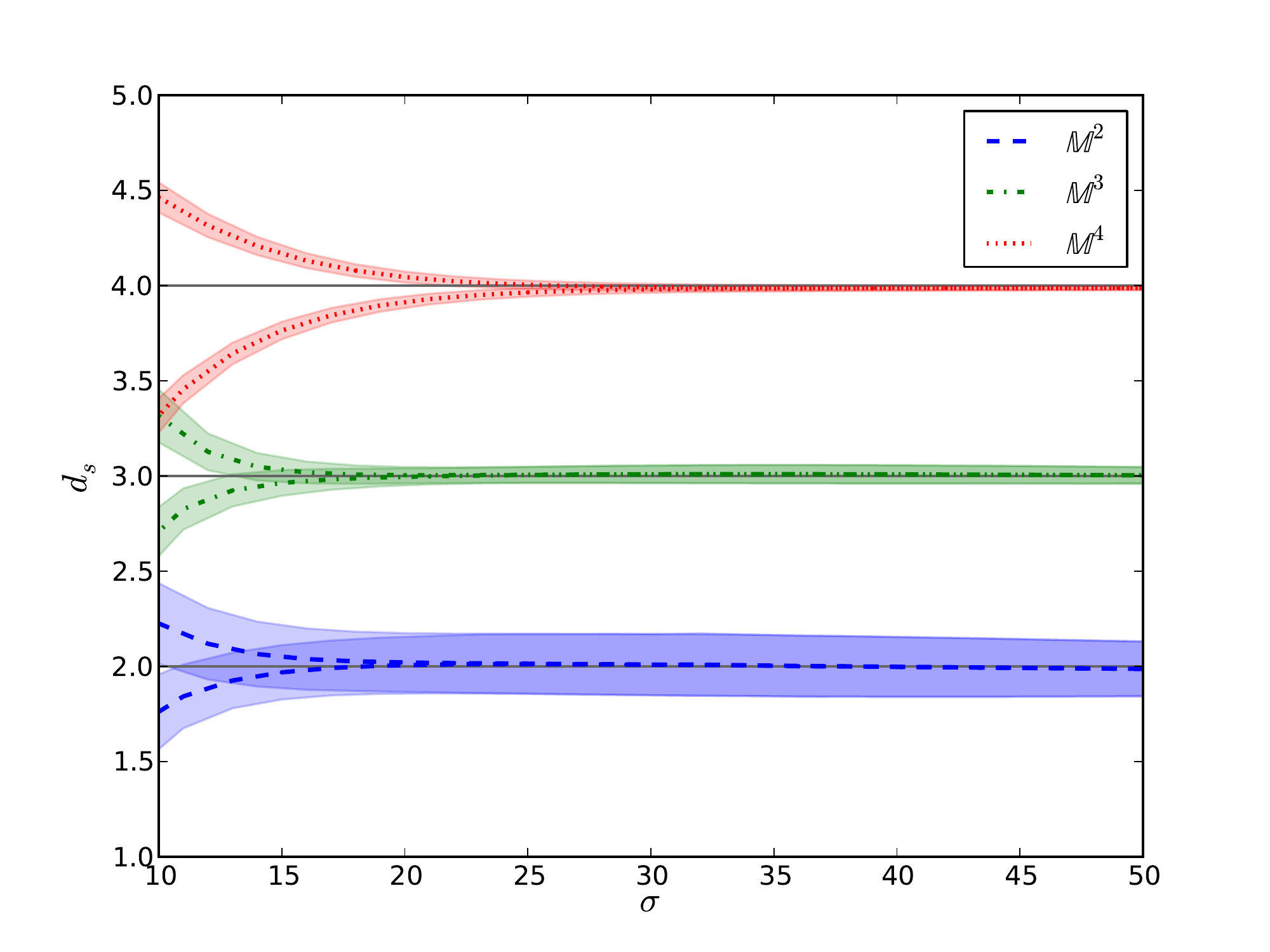}
\caption{\label{m2_dim}Left panel: Spectral dimension for $1+1$ dimensional Minkowski spacetime { ($N \approx 10^8$)} as a function of $\sigma$,  for different values of the cutoff. After initial oscillations the curves for even and odd diffusion steps merge at $\sigma \sim 20$ with $d_s>2$ and decrease with $\sigma$. 
Higher cutoff values yield larger initial values of the spectral dimension. Independently of the cutoff, all the curves approach $d_s = 2$ in the intermediate-$\sigma$ regime, before dropping to 0 at even larger values.\\
Right panel: Spectral dimension for sprinklings into Minkowski spacetimes with different dimensionality. Here we used a low cutoff value of $L=5$ in order to make the computation for higher dimensions possible. The small $\sigma$ regime exhibits the split of the spectral dimension in odd and even time steps.
The spectral dimension 
approaches
the topological dimension in all the cases. { The higher dimensional results are less uncertain, as the effective volume probed becomes larger and fluctuations are less important.} Here we set $L$ constant, so $N \sim L^d$ and hence larger values of $d$ probe larger $N$ which gives a smaller uncertainty as the Poisson noise goes as $\Delta N \sim \sqrt{N}$, so the 
statistical error $\Delta N/N \sim 1/\sqrt{N}$.}
\end{figure}

In our simulations with a finite cutoff,
the intermediate-$\sigma$ regime of the
spectral dimension approaches the topological dimension, cf. fig.~\ref{m2_dim}. 
This is to be expected, as the spectral dimension can be used to test whether a smooth classical spacetime emerges from a quantum gravity setting at large scales, in which case $d_s \rightarrow d$.
As a major difference of the causet to a regular lattice, where the asymptotic value $d_s \rightarrow d$ is approached from below, the small-scale value for the spectral dimension on a sprinkling into Minkowski spacetime is larger than the topological dimension, due to the residual nonlocality of causal sets even in the case with a cutoff, unless we choose a cutoff $L \geq l_{\rm Pl}$, as in the right panel of fig.~\ref{m2_dim}.

In the same way, we can study higher-dimensional Minkowski spacetime, cf. fig.~\ref{m2_dim}. The higher-dimensional cases are clearly limited computationally, since the volume needed to accommodate for the diffusion without boundary effects gets much larger. Hence we are limited to simulations with small cutoff values. Nevertheless, the large-$\sigma$ asymptotic behaviour demonstrates how the spectral dimension tends to the topological one.

\begin{figure}[!h]
\includegraphics[width=0.5\linewidth]{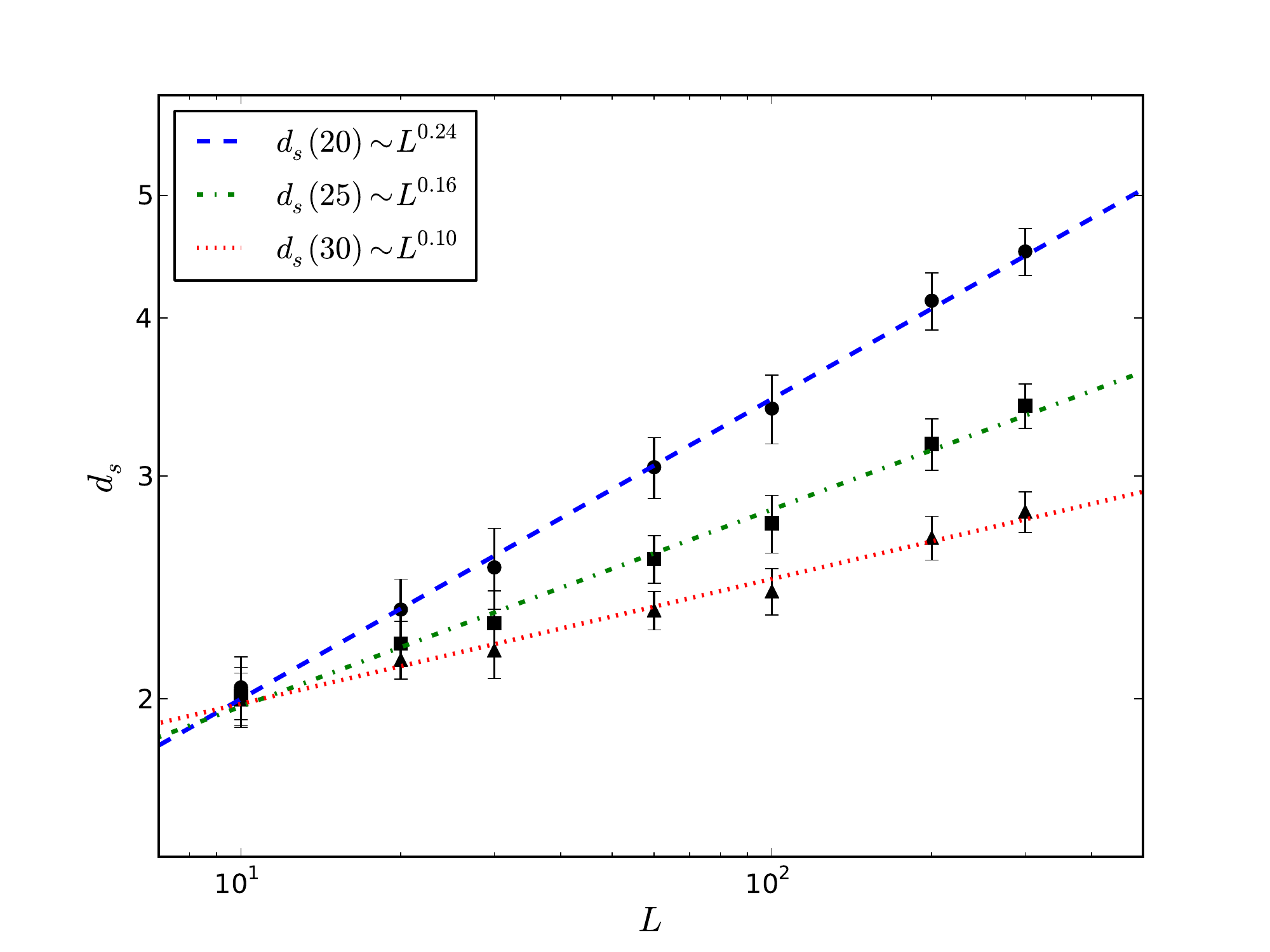}
\caption{\label{m2_d_star}{ Value of $d_s$ measured at $\sigma_\ast=20,25,30$ (right after the even and odd $\sigma$ lines merge) against the cutoff distance $L$ {{for $N \approx 10^8$}}. It is evident that the early-$\sigma$ value follows the power law $d_s(\sigma_\ast=20) \sim L^{0.24}$ and hence will be divergent for $L \rightarrow \infty$. In this limit we 
expect that the large-$\sigma$ behaviour of $d_s$ will 
resemble
the case with a finite cutoff.}}
\end{figure}

Let us consider the spectral dimension as a function of the cutoff and the diffusion time $\sigma$: For very low values of the cutoff, when most links are cut off, the spectral dimension is lower than $d$ for small $\sigma$. 
In this case the cutoff is so severe that it effectively reduces the dimensionality of the causal set. 
For larger values of the cutoff, the causal set non-locality becomes apparent. Its effect can be clearly seen by considering $d_s (\sigma_{\ast})$ as a function of $L/l_{\rm Pl}$. Here we define $\sigma_{\ast}$ as the (approximate) point where the even-$\sigma$ and odd-$\sigma$ curves are already merged. As expected, $d_s (\sigma_{\ast})$ grows for increasing $L/l_{\rm Pl}$. The rate of growth becomes slower if we choose larger values of $\sigma$ where to measure $d_s$, cf. fig.~\ref{m2_d_star}.
The limit of a Lorentz invariant sprinkling into Minkowski spacetime accordingly shows a divergent small-scale spectral dimension, as expected. For a finite causal diamond in Minkowski spacetime, $d_s>d$, with a finite value for $d_s$, holds. The large-scale behaviour indicates that the spectral dimension will indeed yield $d_s=d$ in an intermediate regime. For a finite causet, we expect the drop-off to $d_s=0$ in the limit $\sigma \rightarrow \infty$.

\FloatBarrier
\subsection{Spectral dimension of sprinklings into spacetimes with extra dimensions}

\begin{figure}[!h]
\includegraphics[width=0.48\linewidth]{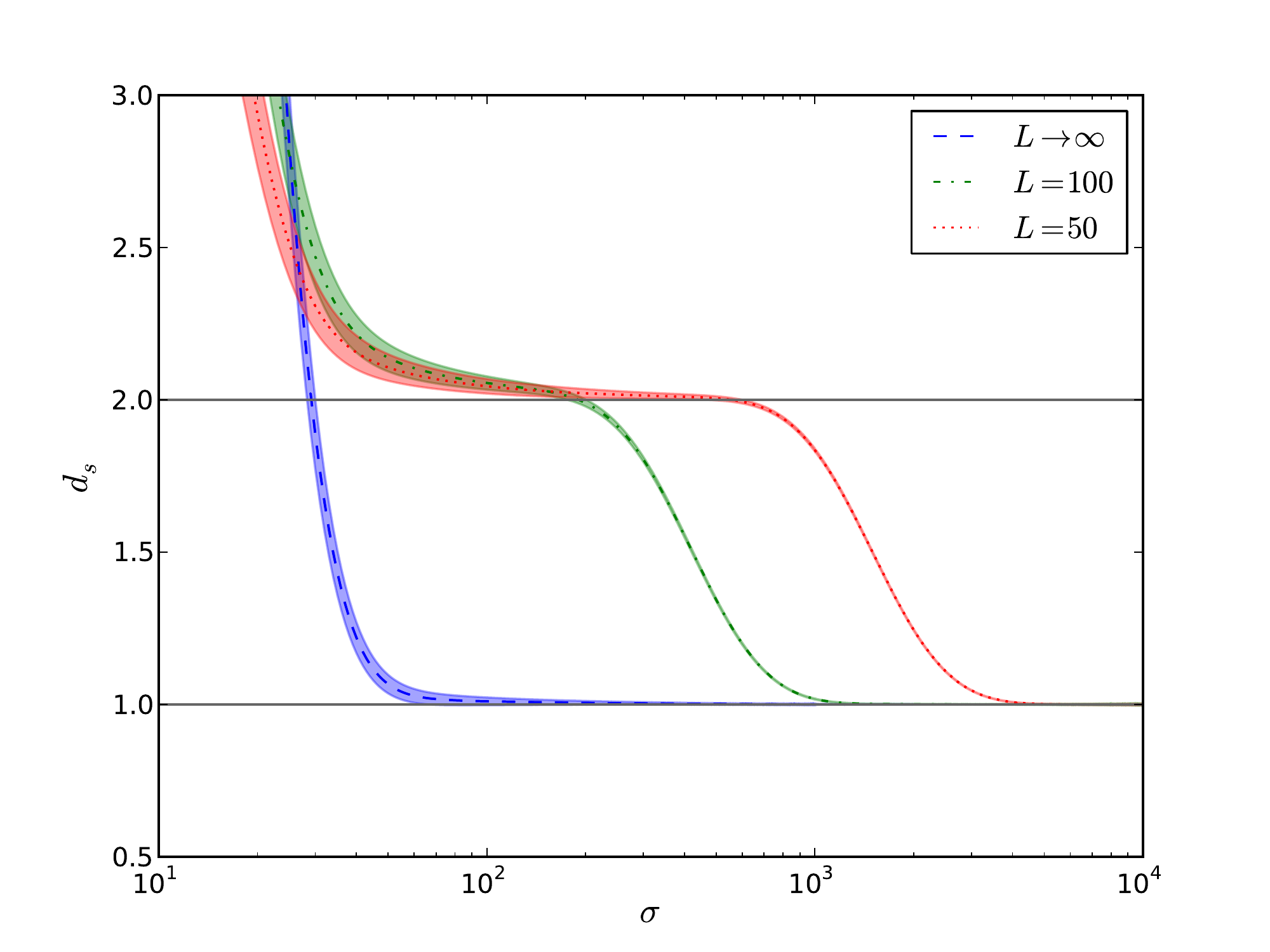} \includegraphics[width=0.48\linewidth]{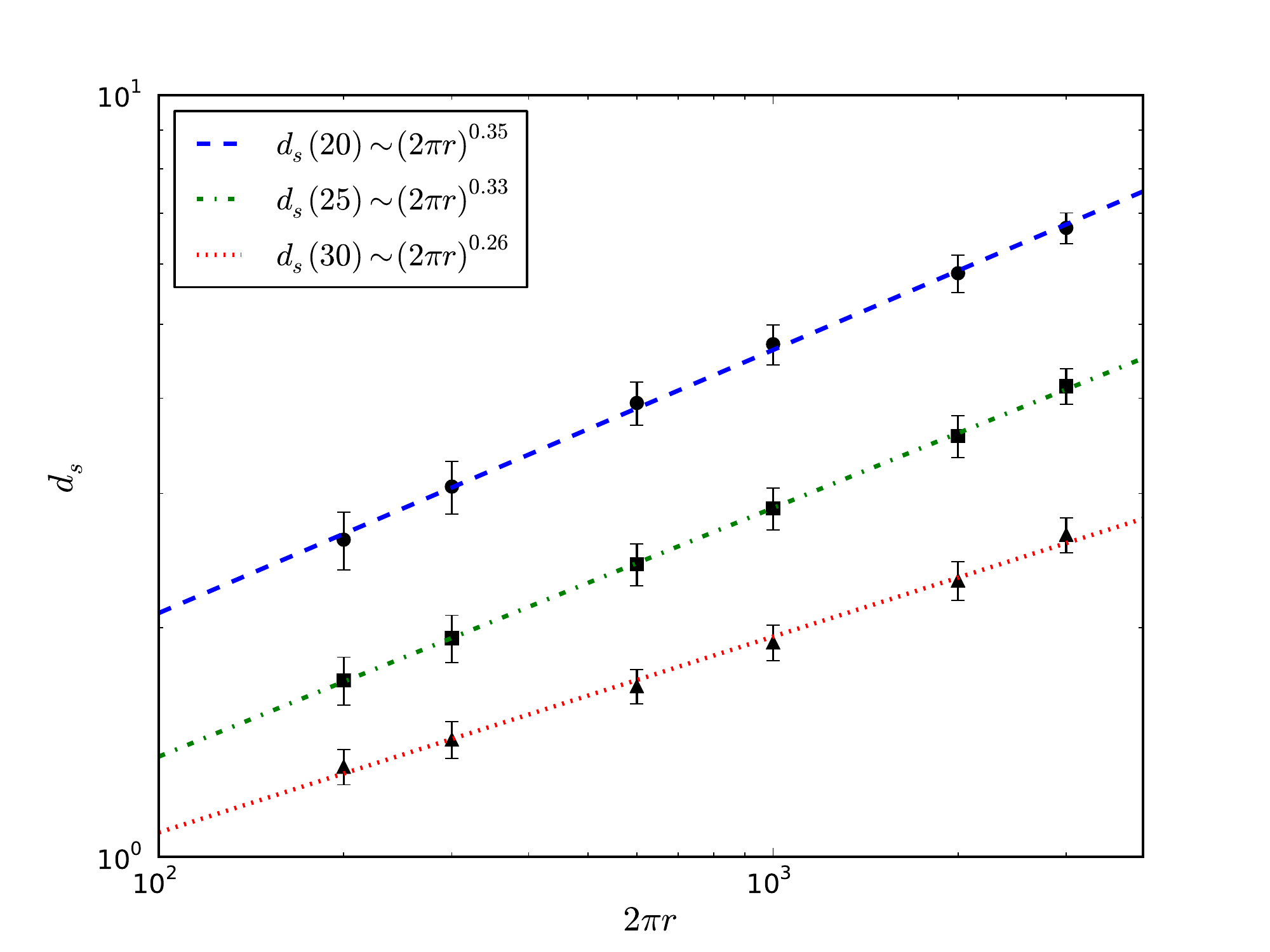}
\caption{\label{cyl2d_dim}Left panel: Spectral dimension in the case of 2d cylinder space with $2 \pi r = 1000$ as a function of diffusion time $\sigma$. We show the spectral dimension for a cutoff $L=50$ (green dot-dashed line) and $L=100$ (red dotted line), and the case without a cutoff (blue dashed line). In all cases, the spectral dimension obeys $d_s\gg 2$ at small $\sigma$. 
The introduction of the cutoff reveals a plateau that becomes more extended with lower values of $L$.\\
Right panel: Spectral dimension in 2d cylinder space evaluated at $\sigma_\ast=20,25,30$ as a function of the compactification radius. We find a fit of $d_s(\sigma_\ast=20) \sim (2\pi r)^{0.35}$ that indicates the small-scale $d_s$ would diverge as we consider larger and larger radii, thus agreeing with the spectral dimension on $1+1$ Minkowski spacetime. Note that we expect the power law fit to hold only for sufficiently large $r$, and hence we neglected $2\pi r < 200$ in the analysis.}
\end{figure}

Let us consider a spacetime with a compact dimension, namely $\mathbb{R} \times S^1$. In contrast to Minkowski spacetime, its causal structure 
implies that each element only has a finite number of links. This follows by considering a point $x$  and integrating the probability that it has a link to some point $x'$, $e^{- \rho V(x,x')}$ over the infinite volume of $\mathbb{R} \times S^1$, which yields a finite number. We thus expect a finite spectral dimension $d_s>d$ at small $\sigma$, even at infinite volume. For $\sigma \gg r$, where $r$ is the compactification radius, we expect an effectively one-dimensional diffusion process.
In our simulations, the spectral dimension reduces to $d_s=1$ at large $\sigma$, in accordance with our expectation, cf. fig.~\ref{cyl2d_dim}.
We expect that for large compactification radius $r \gg l_{\rm pl}$ there should be an intermediate plateau regime, where $d_s=2$, and $d_s=1$ only for larger diffusion times. In this regime, most links do not extend around the whole radius of the cylinder, but the causal set starts to resemble a sprinkling into flat Minkowski spacetime without compactification. 
It turns out that for computational reasons this regime is hard to reach, as it requires very large sprinklings. We therefore use the following trick to mimic that regime: Introducing a cutoff $L$ beyond which we discard links has the effect to remove links which extend very far around the cylinder. Accordingly this approximates the causal set for a larger $r \gg l_{\rm pl}$.
In that case our simulations explicitly show the emergence of the 2-dimensional plateau before going to $d_s=1$, cf. fig.~\ref{cyl2d_dim}.

The small-$\sigma$ value of the spectral dimension $d_s(\sigma_\ast)$ satisfies $d_s>d$, but remains finite. As the limit $r \rightarrow 0$ corresponds to the limiting case of 2$d$ Minkowski spacetime, the spectral dimension should increase as a function of compactification radius $r$. We find that a power-law with $\sigma$ dependent exponent fits the data well, see fig.~\ref{cyl2d_dim}, suggesting that the spectral dimension diverges in the limit of infinite volume and $r \rightarrow \infty$. As the small-$\sigma$ value of the spectral dimension depends on $r$, we observe that the spectral dimension is sensitive to the global topology of the spacetime. Again this is due to the nonlocality of causal sets, which implies that a random walker can access global information such as the total volume or the global topology, already in the first steps of a random walk.

\FloatBarrier
\subsection{Spectral dimension of sprinklings into de Sitter spacetime}\label{deSitter}
\begin{figure}[!here]
\includegraphics[width=0.33\linewidth]{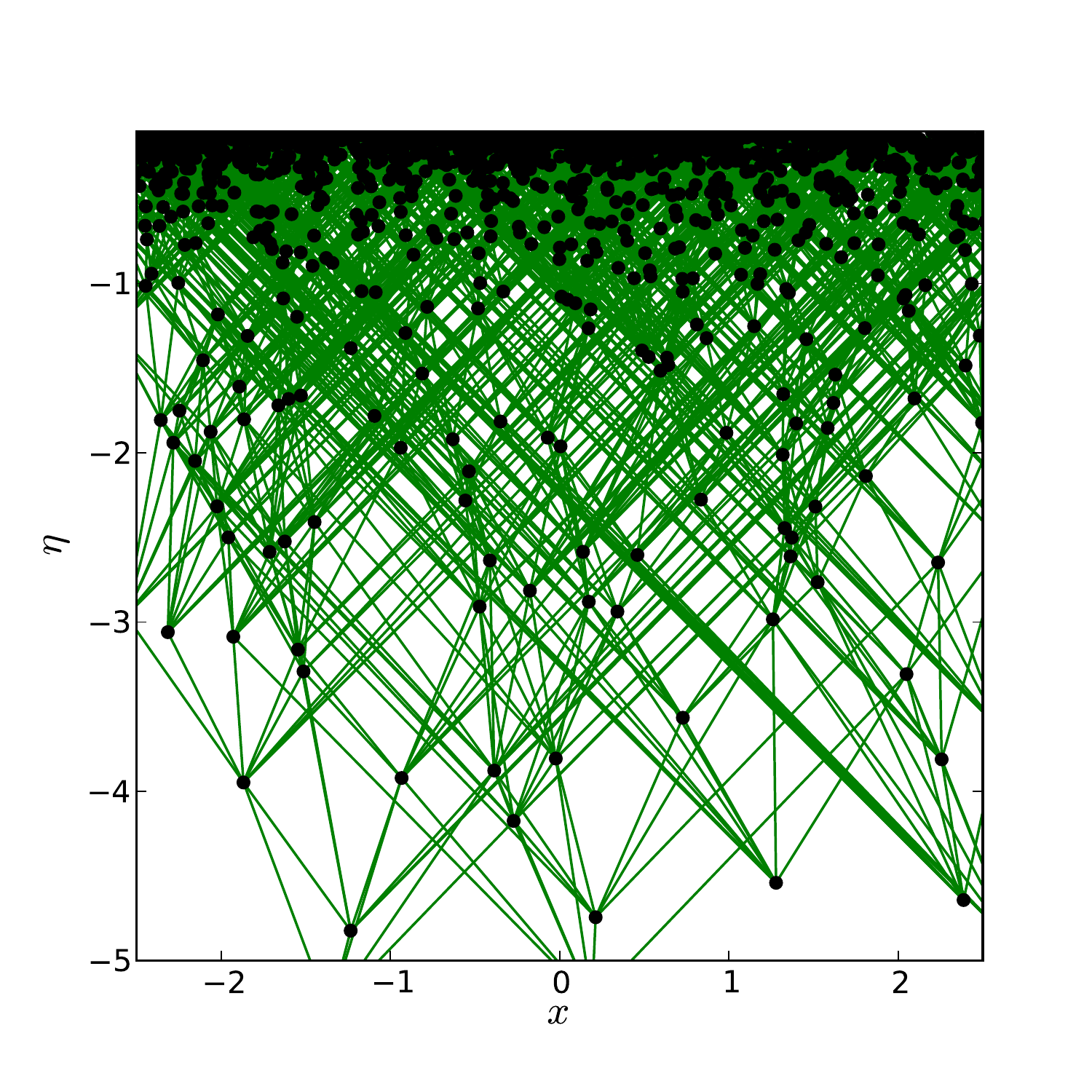}
\includegraphics[width=0.33\linewidth]{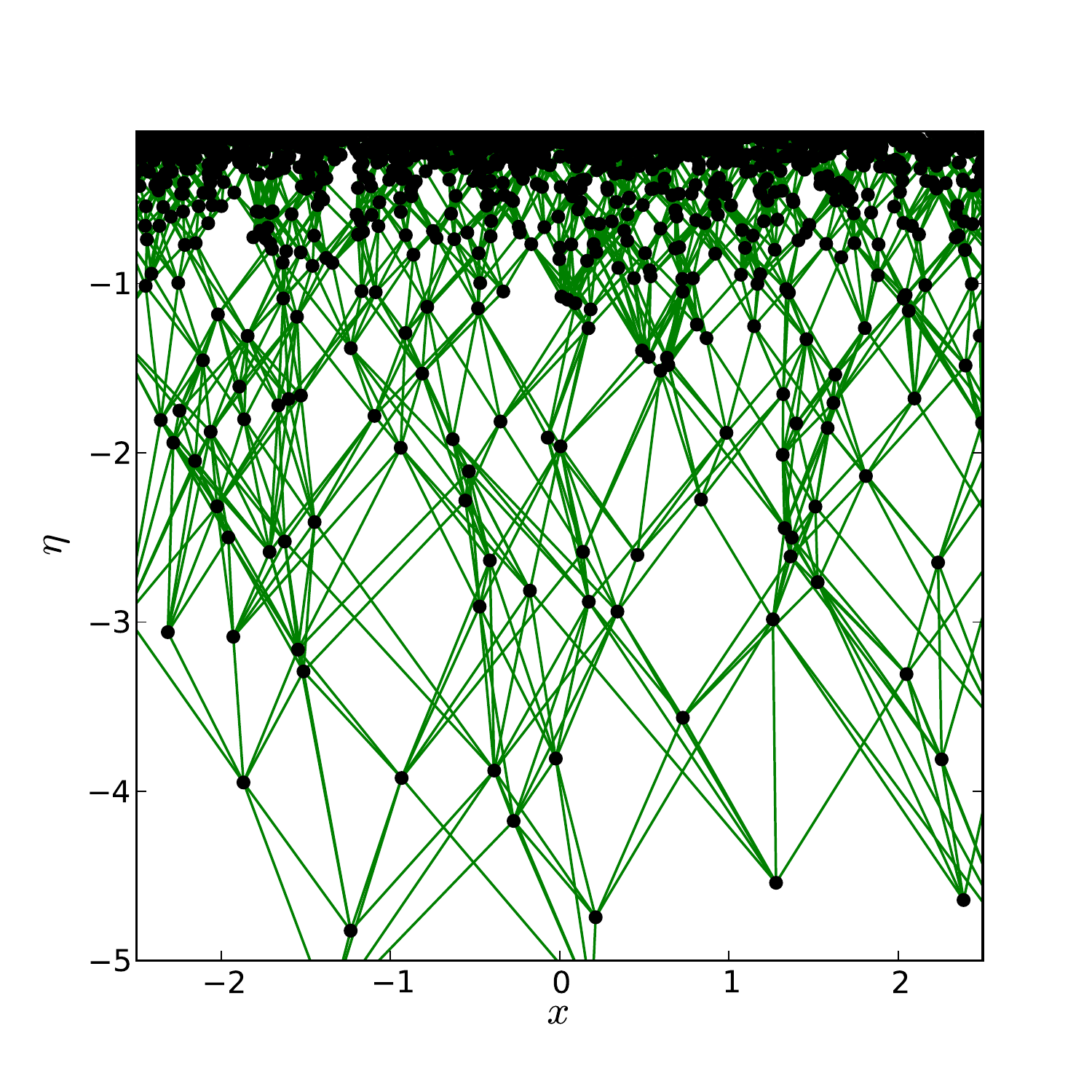}
\caption{\label{sprinkle_ds2}{Sprinklings into $(1+1)$-dimensional de Sitter spacetime (planar patch) with unit density and radius $\alpha=5$: without any cutoff (left), and with infrared cutoff $L^\ast=0.5$ (right).}}
\end{figure}

The spectral dimension of de Sitter spacetime is of particular interest, as its cosmological patch describes 
the inflationary universe and the $\Lambda$ dominated late-time expansion of the universe. 
For the cosmological or planar patch of 1+1 dimensional de Sitter, which covers half the space, namely 
the region inside and including the lightcone emanating from a point on spacelike past infinity $\mathscr{I}^-$, we can use conformal coordinates $(\eta, x)$, in which the 
metric is given by $g_{\mu \nu} = \left(\frac{\alpha}{t}\right)^2 \eta_{\mu\nu}$. Here $\alpha$ is the de Sitter radius and the coordinates have range $t \in [-\infty, 0]$, $x \in [-\infty,\infty]$. 
Creating a sprinkling into this patch works as follows: After sprinkling into a given interval $[x_0,x_1] \times [t_0,t_1]$, the links follow from the causal structure of Minkowski spacetime. To account for the difference to de Sitter, it is only necessary to adapt the sprinkling density: As $\sqrt{-g} = (\alpha/t)^2$, the sprinkling density will not be constant, cf. fig.~\ref{sprinkle_ds2}. Following \cite{BSchmitzer, Aslanbeigi:2013fga} we achieve this by first sprinkling with constant density into a unit square, denoted by $A$ and then using a mapping $f(A) \rightarrow \Omega, f(x) \mapsto z$ for the coordinates, such that $\int_A d^2x = \frac{1}{V(\Omega)} \int_{f(A)} d^2 z \sqrt{-g}$.

To check whether global properties can affect the spectral dimension, we will consider random walks on a sprinkling into the planar patch, as well as a sprinkling into the global patch. The metric for global de Sitter spacetime can, in 2 dimensions, also be brought into a conformally flat form and reads $g_{\mu \nu}= - \frac{\alpha^2}{\cos(t)^2}\eta_{\mu \nu}$, where $t \in [- \pi/2, \pi/2]$. While a causal diamond in the cosmological patch can be mapped onto a causal diamond in the global patch due to the symmetries of de Sitter spacetime, the global properties of both patches differ \cite{Aslanbeigi:2013fga}. 
Nevertheless we do not observe any qualitative differences of the spectral dimension between the two patches in our simulations, hence in the following discussion we will consider the planar patch only.

In the cosmological patch, the number of past-directed links is finite at every point in a causal interval \cite{BSchmitzer}.
As the number of future-directed links from a given point will however not be finite unless we restrict ourselves to a finite volume, we again resort to a regularisation to evaluate the spectral dimension.
Following our previous considerations, we introduce an infrared cutoff, again measuring the distance of points after a Wick-rotation. The cutoff is defined as $L(\eta) = \rho_0 \eta L^\ast$, which is position-dependent to account for the changing density by keeping the number of future-directed links from each element constant, cf. fig.~\ref{sprinkle_ds2}.

\begin{figure}[!h]
\includegraphics[width=0.48\linewidth]{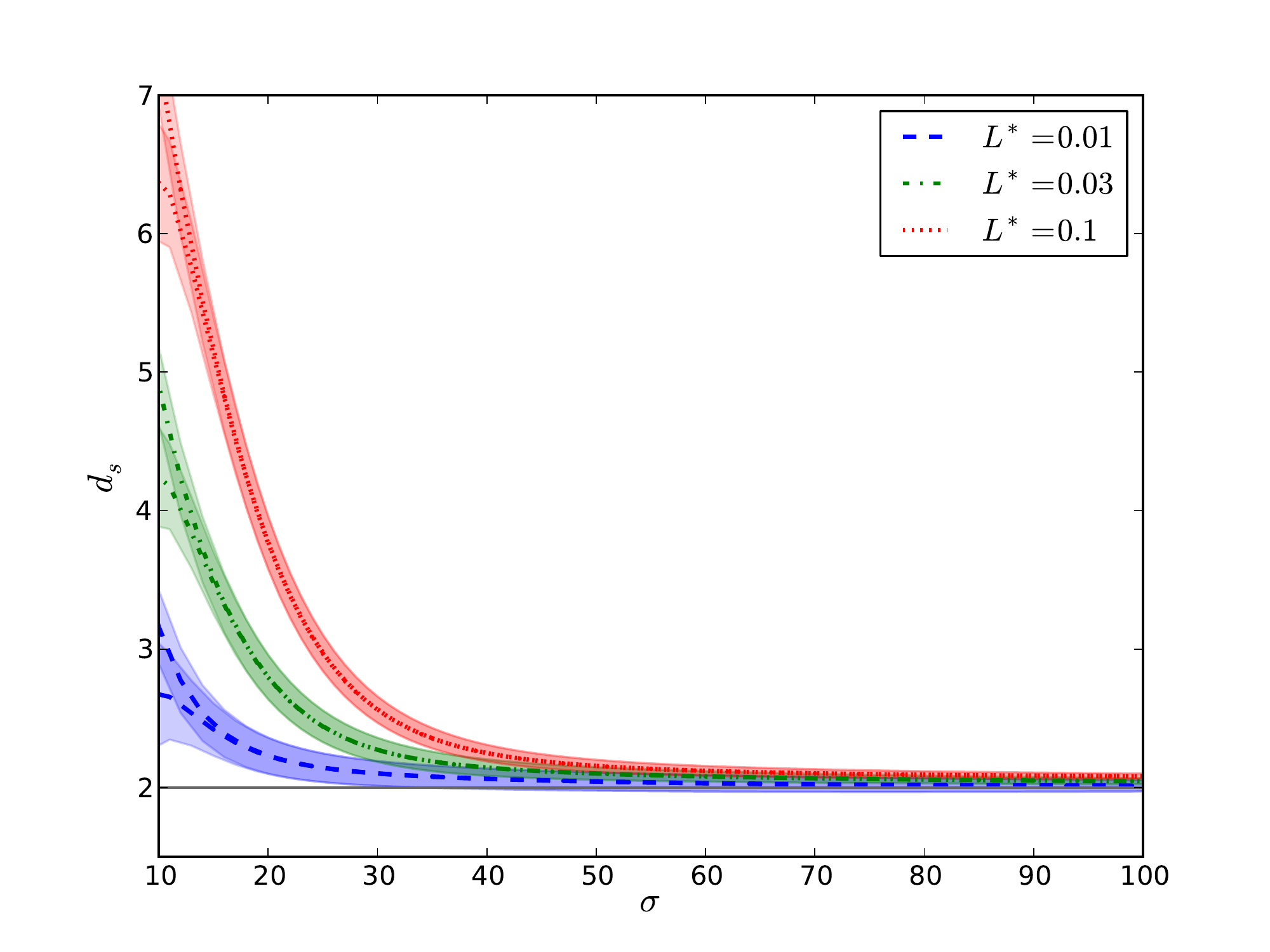}
\includegraphics[width=0.48\linewidth]{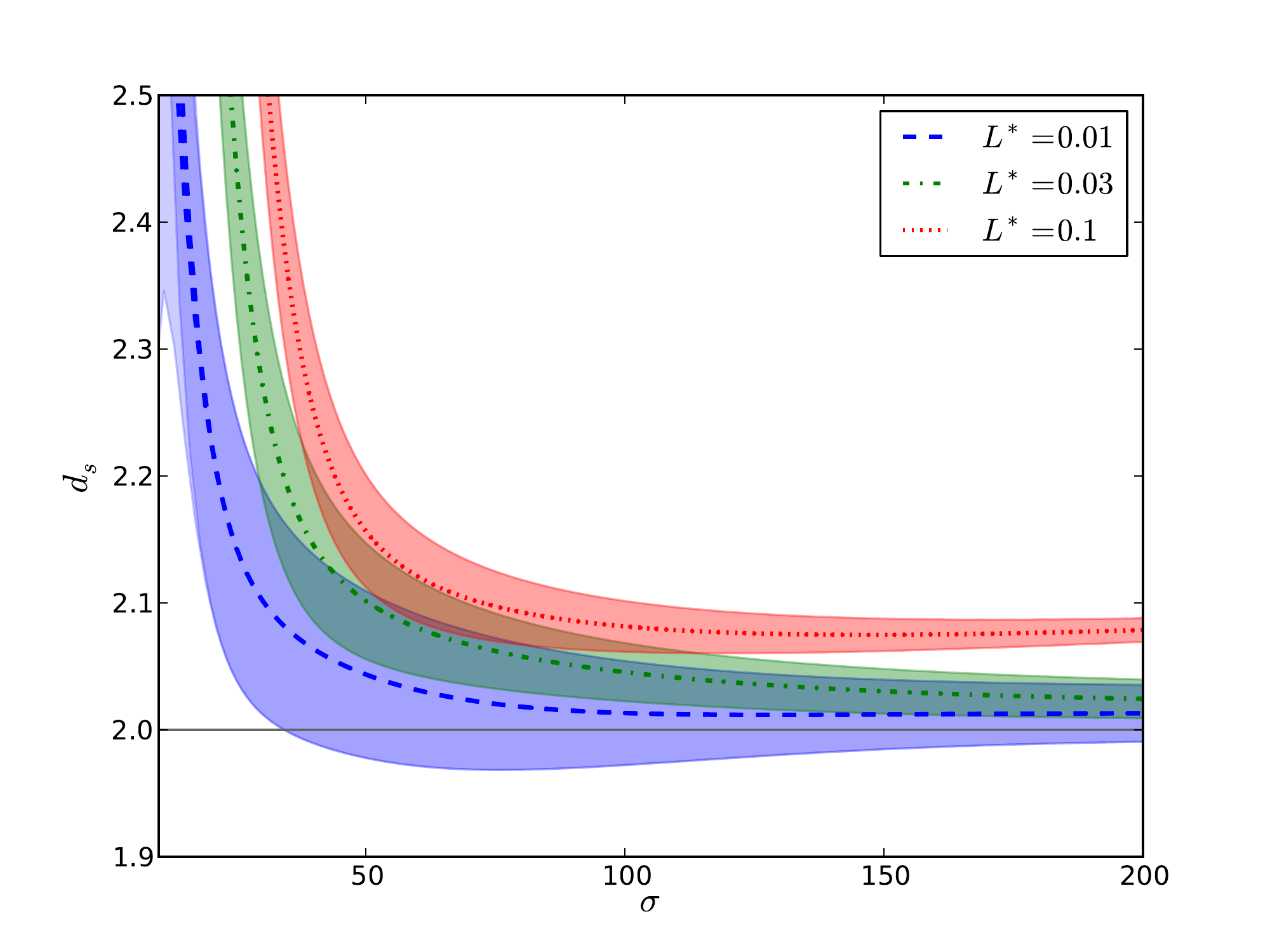}
\caption{\label{ds2_dim} Spectral dimension $d_s$ as a function of the diffusion time $\sigma$ in $1+1$ de Sitter spacetime (planar patch) with radius $\alpha=10^3$, starting point $(\eta,x) = (1,0)$, and three different cutoff scales $L^\ast = 0.01, 0.03, 0.1$. Similar to the diffusion on a flat background we observe an early increase of the spectral dimension on small scales, however on large scales its value increases compared to the topological dimension of $2$ which we tentatively attribute to the presence of curvature. We observe no qualitative differences to the global patch.}
\end{figure}

If we choose a curvature scale that is considerably larger than the fundamental length, the small $\sigma$ behaviour of the spectral dimension should approach that of Minkowski spacetime.
Accordingly, we can observe the spectral dimension to take a 
value $d_s>d$ at small $\sigma$
and then approach the intermediate value $d_s=d$. 
The large $\sigma$ behaviour will be affected by the presence of curvature.
We tentatively conjecture the late-$\sigma$ increase to be due to the presence of the curvature, cf. fig.~\ref{ds2_dim}. Choosing a larger value of the cutoff implies that the random walker is affected by the curvature at earlier diffusion times.

\FloatBarrier
\subsection{Spectral dimension of KR orders}
\begin{figure}[!here]
\includegraphics[width=0.15\linewidth]{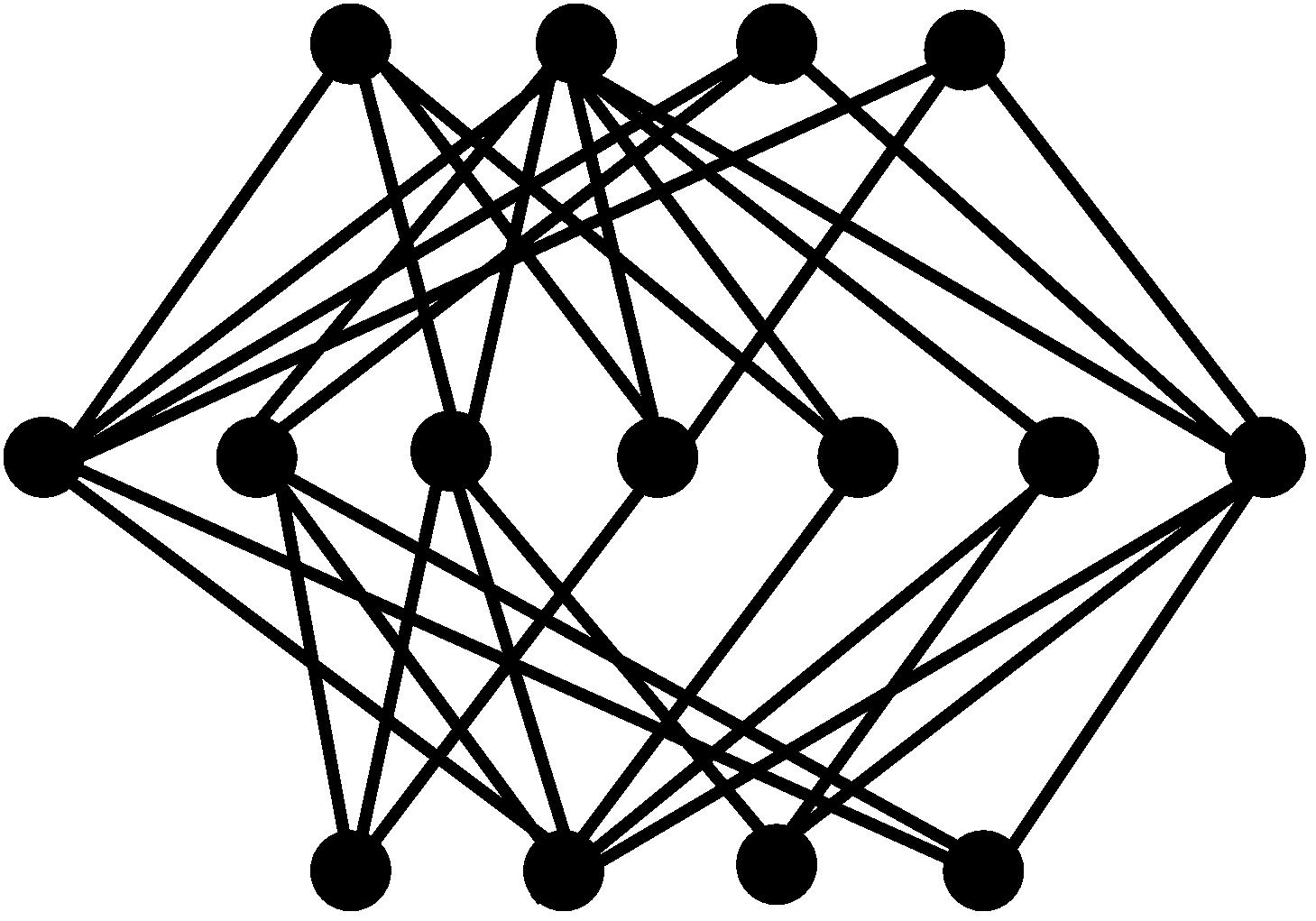}
\caption{\label{KRorder}Example of a KR-type order for small $N$.}
\end{figure}

Kleitman-Rothschild (KR) orders are an important subclass of all causets with $N$ elements: They consist of three layers of elements, with approximately $ N/2$ elements in the middle layer, and approximately $ N/4$ elements in the upper and lower layer. Each element in the middle layer has probability $1/2$ to share a link with any given element in the upper and lower layer. There are no direct links between the lower and the upper layer, nor within each of the layers, see fig.~\ref{KRorder}. In the limit of large $N$, KR orders are entropically favored, since their number grows as $e^{N^2/4}$, similar to the total number of causal sets of order $N$, which also grows as $e^{N^2/4}$ at leading order \cite{Kleitman}. Due to their structure, KR orders do not approximate manifolds \cite{Rideout:1999ub}, as they only contain three discrete ``moments in time".

To test whether the spectral dimension reflects the non-manifoldlikeness, let us consider a random walk averaged over KR orders: Starting from any given point, the first step takes us into an element in the neighbouring layer. The second step can bring us back to the starting point, with a probability of $P(x,x,2)= 2/N$ ($N>4$), when averaged over KR orders. This probability is constant in the three layers, as it is the probability for the existence of a given link ($1/2$), multiplied by the average number of links, which is $N/4$ for all three layers. This pattern continues for arbitrarily large $\sigma$: 
\be
P(x,x,\sigma) = \begin{cases}
\frac{2}{N}, &\quad \sigma = 2k\\
0, &\quad \sigma =2k+1.
\end{cases}
\ee
 This oscillating behaviour is a manifestation of a discrete setting. In the case of KR orders, it is present for arbitrarily large diffusion times $\sigma$,
signaling the non-manifoldlikeness of this class of causets.

Let us introduce a finite probability $\chi<1$ that the diffusion particle remains at the same point. This smoothens the odd- and even-$\sigma$ curves, and changes the early-$\sigma$ behaviour of the return probability (which decays as $\chi^\sigma$), however its asymptotic value still converges to a constant, and hence the spectral dimension vanishes on large scales. By explicitly simulating KR orders we confirmed this fact, and also observed that larger orders yield a faster decay of $d_s$.

\subsection{Spectral dimension of transitive percolation models}

Much work has been done to construct dynamics for causal sets, which lead to the development of so-called Classical Sequential Growth models \cite{Rideout:1999ub, Rideout:2001kv}. Of particular interest is a specific subclass of these, namely the transitive percolation model \cite{Rideout:2000fh}.
There, a causal set is ``grown'' successively in a label-invariant way, which is the remainder of diffeomorphism invariance in the discrete setting: Given an $N$ element causal set, a new element will be included. The probability for it to have a link with any element in the existing causal set is fixed to a small, non-vanishing value $p$. These models are attractive as they define an intrinsic (albeit classical) dynamics, and can be shown not to contain KR orders, as  the probability of obtaining a chain of length greater than $3$ is always non-zero \cite{Rideout:1999ub}. 

 Further modifications to the model aiming at obtaining a $(3+1)$-dimensional continuum limit have also been studied \cite{Rideout:2000fh}. The causal set constructed in this way contains an infinite number of so-called posts, i.e. single elements between which the universe cycles through phases of expansion, stasis and contraction \cite{Bollobas2}, and it has been suggested that the early expansion phase after the post may resemble $(3+1)$-dimensional de Sitter spacetime \cite{Ahmed:2009qm}. The resulting causet has a Myrheim-Meyer dimension of one \cite{Bollobas1}.

Let us now consider a random walk averaged over causal sets of cardinality $N$ grown in this way. Starting from any given point, the probability for a link to any other point is $p$. Thus, the probability for the random walker to return to the initial point is always $p/(Np) = N^{-1}$, independent of the diffusion time and position. Accordingly the spectral dimension will be zero for this class of models. 

If we consider only a single transitive percolation, instead of averaging over the whole class, the large-$\sigma$ limit obtained from averaging over the starting point will also be $d_s=0$, 
using the same argument. 
We can conclude that the percolation model does not give a finite spectral dimension, which supports the point that the resulting causal sets are not manifold-like. 

This has implications for the path-integral over causets: As for large $\sigma$, $d_s=0$ for transitive percolation models and KR orders, these do not yield a contribution to $\langle  d_s \rangle$, irrespective of the microscopic dynamics. This suggests that the expectation value of observables such as $d_s$ could actually show a large-scale behaviour that approximates the expected semiclassical result.

Furthermore, this is an interesting case that illustrates how the spectral and the Myrheim-Meyer dimension can differ.
Our findings on KR orders and transitive percolation models suggest that the spectral dimension could be used as a measure of manifold-likeness, as it vanishes on causets which do not approximate manifolds.

\section{New probe of quantum spacetimes: Causal spectral dimension}\label{meetdim}

In the standard setup that we have considered so far, the diffusing particle explicitly violates causality, as it can propagate 
forwards and backwards in time. In order to introduce a probe of quantum geometry that is closer to what a physical probe, i.e. a physical particle propagating on  a quantum spacetime (neglecting backreaction) could be, 
we introduce a new dimensional estimator. It is based on a random walk that respects the time-ordering of spacetime and respects causality by only propagating within the local lightcone. It is thus applicable to Lorentzian quantum  gravity models and we will study it for causal sets here.
While we so far have treated the causal set as a given graph, on which a probe particle diffuses, we will focus on the interpretation of the links as causal links in this section, imposing causality on the random walks and restricting the movement of the random walker to be forward in time. 
Then the return probability to the starting point is clearly zero, making the spectral dimension ill-defined.   
We therefore consider two causal random walks, and extract a notion of dimensionality from the probability that the two random walks meet at a given diffusion time $\sigma$. 

Let us first consider this process on a flat $d+1$ dimensional space and show how to extract the dimensionality from the meeting probability.
As we impose causality, the movement in the time-direction is restricted to be forward in time and within the local lightcone. The stochastic process underlying such a situation is well-known, and corresponds to a biased random walk, i.e. the probability to move forward and backward in time are no longer equal. {Using light-cone coordinates, we can straightforwardly set up the discrete approximation to this process on a lattice, and then take the continuum limit by taking the lattice spacing to zero. 
We thereby arrive at the drift-diffusion equation in $d+1$ dimensions, where the drift term enforces causality at large diffusion times:
\be
\partial_{\sigma} P(\vec{x},t,\sigma) = - v \partial_t P(\vec{x},t,\sigma) +D_x \nabla_{\vec{x}}^2 P(\vec{x}, t, \sigma) + D_t \partial_t^2 P(\vec{x}, t, \sigma).
\ee
Herein $t$ is the Wick-rotated time coordinate, $v$ is the drift velocity, and $D_x$ and $D_t$ are diffusion constants which can be set to 1 by a choice of units. This equation is also known as a special case of the Smoluchowski equation, which describes, e.g. the diffusion process of charged particles under the influence of an electric field. The solution for $D_x=D_t=1$ and initial condition $P(\vec{x}, t ,0)= \delta^d(\vec{x})\delta(t)$ is then given by \cite{Montroll}
\be
P(\vec{x},t,\sigma) = \frac{1}{(4 \pi \sigma)^{\frac{d+1}{2}}}e^{-\frac{\vec{x}^2}{4 \sigma}} e^{-\frac{(t-v \sigma)^2}{4 \sigma}},\label{biasedrandomwalk}
\ee
which respects causality at large diffusion times (Alternatively, one can consider the Ornstein-Uhlenbeck process, which realises diffusion in the Lorentzian case by a diffusion process in momentum space, from which the spacetime-process is inferred, see, e.g. \cite{Debbasch}. The large-$\sigma$ limit then agrees with \Eqref{biasedrandomwalk}).
Then the probability for two random walkers to meet at some point after a diffusion time $\sigma$ is $P_{\rm meet}(\sigma) = \int dt \int d^d x\, P(\vec{x}, t, \sigma)^2 \sim \sigma^{\frac{-(d+1)}{2}}$. Accordingly we can define a causal spectral dimension
\be
d_{c\, s}(\sigma) = - 2 \frac{\partial \ln P_{\rm meet}(\sigma)}{\partial \ln \sigma},
\ee
for which $d_{c\, s} (\sigma) =d+1$ in the case of a flat background in analogy to the spectral dimension. 
We thus propose $d_{c\, s}$ as a useful novel dimension estimator in quantum gravity models. As it has a built-in causality property, we conjecture that it can differ from the usual spectral dimension, and provides an interesting probe of quantum geometry that could also be worth studying in other approaches to quantum gravity.
In particular, the causal spectral dimension should be sensitive to nontrivial structure in the time-direction, such as a time-dependent background, or the existence of a preferred foliation.

\subsection{Causal spectral dimension on causal sets}
On the Hasse diagram representing the causet, the causal random walk is straightforward to implement: As links only exist for causal relations, a random walker following the links automatically stays within the local lightcones. The only remaining restriction is to observe the directionality of the links.
As in the case of the spectral dimension, we introduce an infrared cutoff in the case of $1+1$ dimensional Minkowski, de Sitter, and cylinder spacetimes, cf. fig.~\ref{meet_all_dim}. In all three cases the causal spectral dimension turns out to agree with the topological dimension for large $\sigma$. This provides evidence that the causal spectral dimension is a useful novel dimension estimator.

\begin{figure}[!here]
\includegraphics[width=0.49\linewidth]{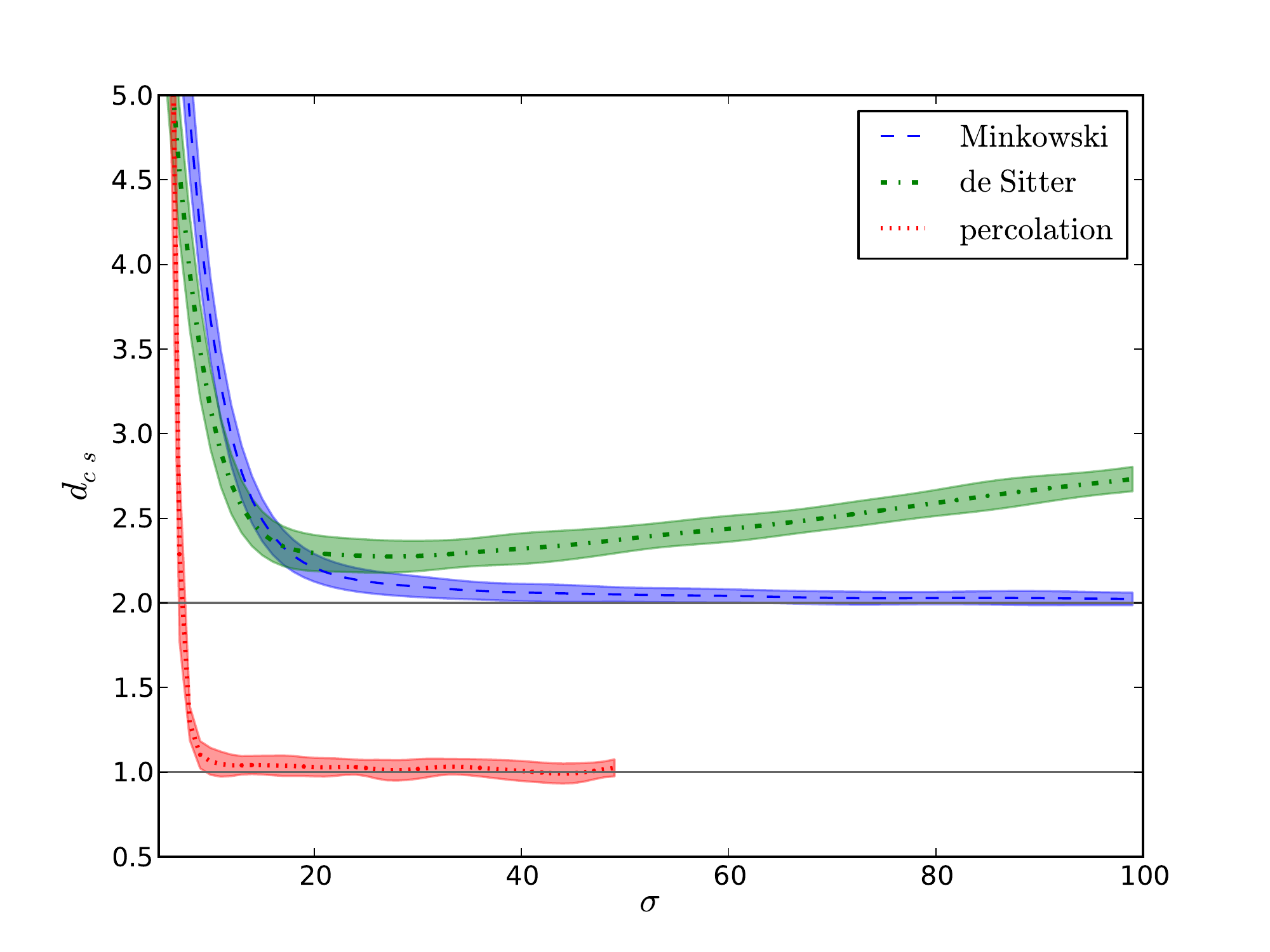}
\includegraphics[width=0.49\linewidth]{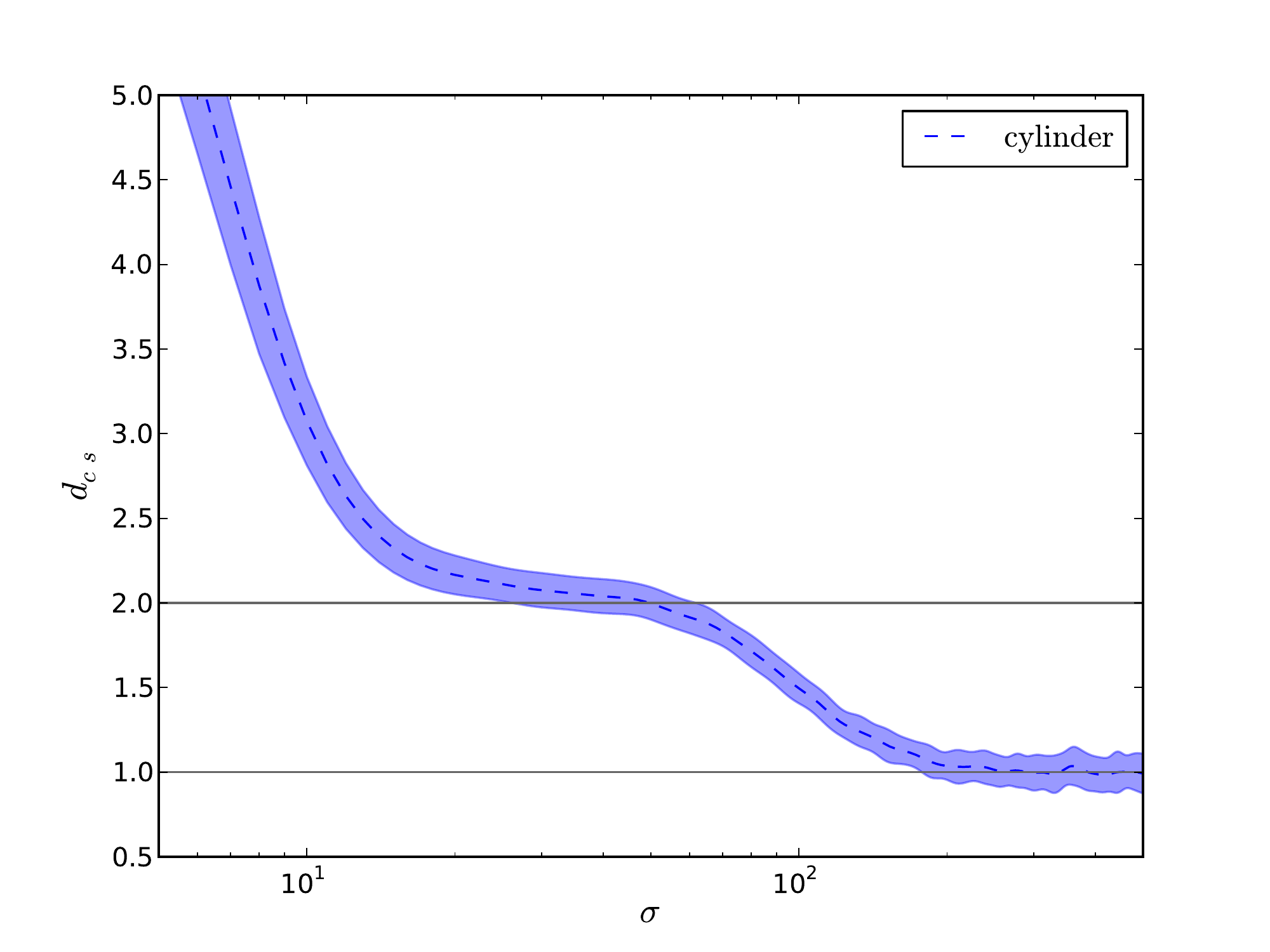}
\caption{\label{meet_all_dim} Causal spectral dimension $d_{c\,s}$ for different causal sets: On the left, sprinklings into $1+1$ dimensional Minkowski ($L=100$, blue dashed line), de Sitter ($\alpha/l_\mathrm{Pl} = 10^3, L^\ast = 0.003$, green dot-dashed line), as well as the transitive percolation model ($N=10^6, p=0.0003$, red dotted line); on the right, cylinder spacetime ($2 \pi r = 300, L=40$). The results for the first three agree with the behaviour of spectral dimension. In contrast, the transitive percolation model yields a non-vanishing asymptotic value of $d_{c\,s} = 1$, whereas $d_s=0$ for this model. We observe that the causal spectral dimension requires up to $10^4$ averages over sprinklings and starting points, since the uncertainties in single simulations become bigger compared to the spectral dimension. Additionally, $d_{c\,s}$ at very low values of $\sigma$ carries significant errors due to calculation of the logarithmic derivative and hence is not plotted.}
\end{figure}

We next consider non-manifoldlike causets, where we find interesting differences to the spectral dimension.
The causal spectral dimension of KR orders is not well defined, as the propagation of diffusing particles is limited to just three time-steps.  
We can study the behaviour of $d_{c\,s}$ on transitive percolation models by performing simulations, where we grow a causet of $N=10^6$ elements. The results indicate that after an initial increase in the causal spectral dimension it quickly converges to $d_{c\,s}=1$ on larger scales, thus agreeing with Myrheim-Meyer dimension \cite{Bollobas1}, in contrast to the spectral dimension, cf. fig.~\ref{meet_all_dim}. Due to computational limitations, we cannot study much larger causets, on which an intermediate regime with $d_{c\, s} \approx 2$ could in principle emerge.
A scale-dependent dimensionality with a large-scale dimension of one in these models has previously been discussed in \cite{Rideout:2001kv, Brightwell}.

\begin{figure}[!here]
\begin{minipage}{0.5\linewidth}
\includegraphics[width=\linewidth]{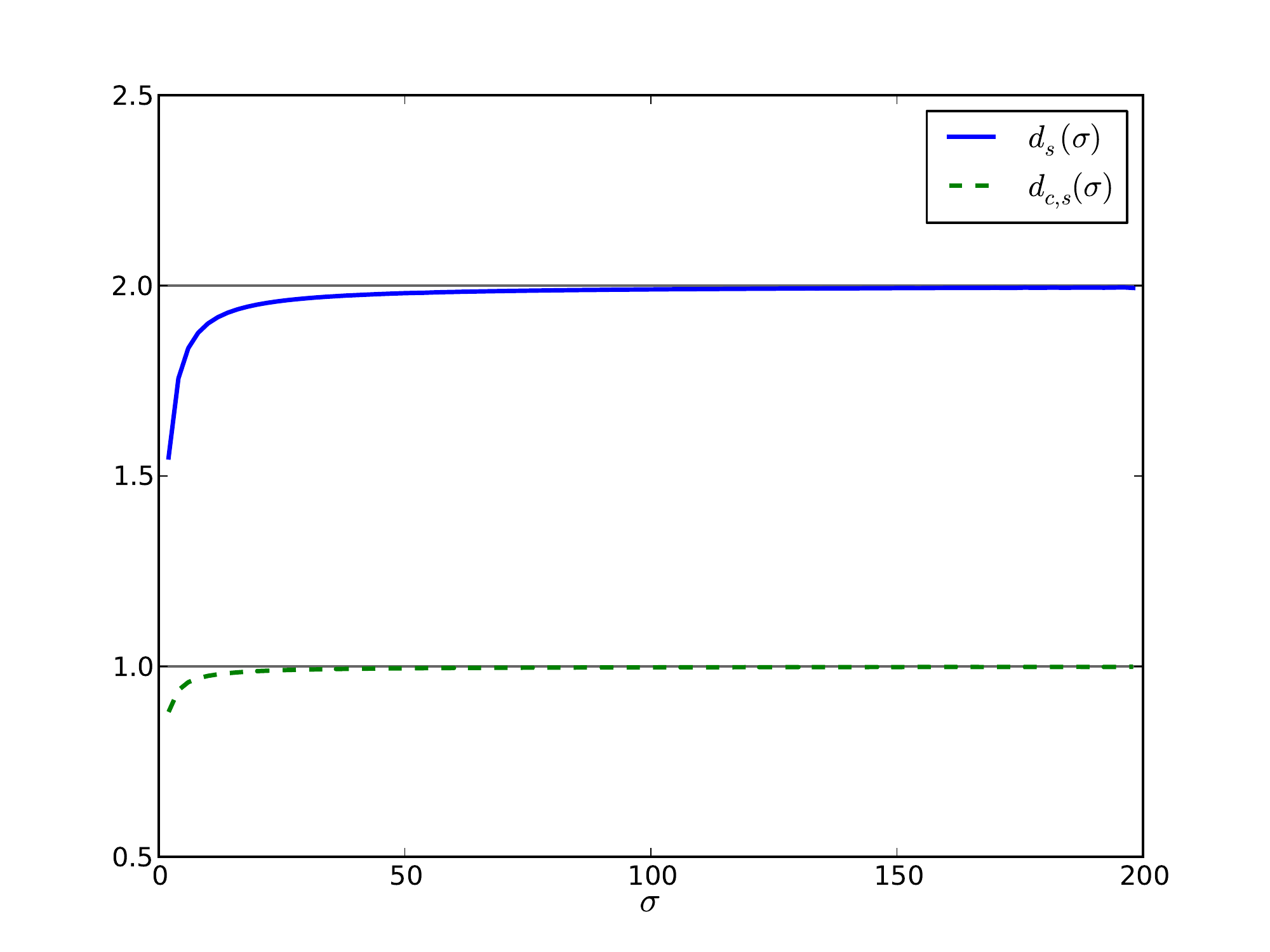}
\end{minipage}
\begin{minipage}{0.48\linewidth}
\caption{\label{tower_dim}Spectral dimension $d_s(\sigma)$ (blue solid line) and causal spectral dimension $d_{c\,s}(\sigma)$ (green dashed line) for a crystalline causet as functions of diffusion time $\sigma$. Due to its regular shape it was enough to perform a single simulation with $N=10^6$ causet elements. We set the size to be larger than the extent of the diffusion to avoid effects of cyclicity of $S^1$, which has been studied separately. Only the even-$\sigma$ curve of spectral dimension is plotted, since all the odd-$\sigma$ values of $P_{meet}$ vanish. After a region of reduced dimensionality for small diffusion times the dimensional estimators converge to: $d_s=2$, $d_{c\,s}=1$. This illustrates how the two can in principle give different results.}
\end{minipage}
\end{figure}

Let us now consider another interesting class of causets, namely crystalline causets, a.k.a. ``tower of crowns'', which are constructed as follows \cite{Major:2009cw}: Taking $m$ antichains (i.e. causal set elements with no links), in which the elements are labeled by $i$, then the $i$th element in the $m$th layer precedes the elements $i$ and $i+1$ mod $m$ in the $m+1$st layer. This yields a regular-lattice-like structure that embeds into $\mathbb{R} \times S^1$. This structure clearly exhibits a preferred time-foliation and is therefore not Lorentz invariant. Interestingly, in this setting the spectral and the causal spectral dimension differ. The reason is simple to understand: The allowed moves for the causal random walk are to hop from element $(m,i)$ to $(m+1,i)$ or $(m+1,i+1)$. There is a vanishing probability to remain in the same layer, or to jump over a layer. Accordingly there is no smearing of the probability density in $m$, it will simply be $\sim \delta(m-\sigma)$ in that direction. This implies, that the causal random walk is effectively only one-dimensional. Accordingly we expect to find $d_{c\, s}=1$ in contrast to $d_s=2$. We confirm this expectation by explicitly simulations of $d_s$ and $d_{c\, s}$, see fig.~\ref{tower_dim}.

This example explicitly confirms, that the causal spectral dimension and the spectral dimension can differ and are independent of each other. In particular, the value of the causal spectral dimension for the crystalline causets is clearly related to the existence of a preferred time foliation, and the corresponding restriction in the allowed steps. We speculate, that in quantum gravity settings with a preferred time foliation such a difference between the causal spectral dimension and the spectral dimension could be common.

It would be interesting to investigate the causal spectral dimension also in other settings, such as, e.g. CDTs. As the local lightcones are well-defined in every building block of a triangulation, causal diffusion can be implemented. This could provide further interesting insight into the microscopic structure of spacetime in CDTs. In particular, it would be interesting to find whether the causal spectral dimension differs when no preferred time foliation is present in that setting, as in \cite{Jordan:2013awa}.

\section{Conclusions}\label{conclusion}
We have studied the spectral dimension $d_s$ in causal set quantum gravity. This probe of quantum spacetimes has received considerable interest, as the small-scale value in several different approaches to quantum gravity has been shown to be equal, and also smaller than the topological dimension $d$. Here, we derive the spectral dimension for causal sets by simulating random walks on the Hasse diagram defining a causal set. The causal set approach to quantum gravity combines Lorentz invariance with discreteness, and posits that the fundamental quantum nature of spacetime is a causal set, i.e. a set of elements (spacetime points), with a partial order (causal relations), which is locally finite, and therefore kinematically discrete. The combination of Lorentz invariance with discreteness implies nonlocality, as elements at large spatial distance -- in a given frame -- will be connected by links.
Our study focuses on the kinematical level, where we consider causal sets that are approximated by manifolds ($\mathds{M}^2, \mathds{M}^3, \mathds{M}^4, \mathds{R} \times S^1, 1+1 \mathrm{\ de\ Sitter}$).  We find that, in contrast to other approaches to quantum gravity, causal sets do not show a dynamical dimensional reduction at small scales. Instead, the spectral dimension shows an increase at small scales, $d_s>d$. We link this behaviour to the nonlocality in causal sets: A random walker has access to a large part of a causal set already within a few time-steps, in contrast to a local setting, yielding a large spectral dimension. We tentatively suggest that such an increasing spectral dimension could occur in other Lorentzian quantum gravity approaches: Whenever the proper distance is used to define a notion of nearest neighbours, an infinite number of nearest neighbours is to be expected in a spacetime of infinite volume. Then some of those nearest neighbours lie at arbitrarily large spatial distances in a given frame, and the diffusion process will exhibit superdiffusion and an increased small-scale value of the spectral dimension.

Furthermore, the nonlocality implies that global information, e.g. on the total spacetime volume, is stored in the number of nearest neighbours of an element. Accordingly the small-scale value of the spectral dimension is sensitive to global properties of the quantum spacetime, and increases with the total volume, e.g. in the case of causal sets that are approximated by Minkowski spacetime.

We observe that, using an infrared cutoff that we impose on the maximal length of a link using the Wick-rotated metric and thus selecting a preferred frame to make the evaluation computationally feasible, the large-scale spectral dimension approaches the topological dimension for manifoldlike causal sets. This is similar to the case of lattice-like discrete approximations of a continuum spacetime, where the spectral dimension equals the topological dimension for large diffusion times.
We further observe that the spectral dimension approaches zero for large diffusion times in all cases of non-manifoldlike causal sets that we have considered (KR orders, transitive percolation models). On the basis of these two observations we conjecture that the spectral dimension is a useful dimensional estimator in causal set quantum gravity, and can be used as a measure of manifoldlikeness. 

We conjecture that an increase, instead of a reduction, of the spectral dimension at small scales could be common to approaches to quantum gravity which are nonlocal, although the precise form of the nonlocality is probably important \cite{Modesto:2011kw}. It would be interesting to infer properties of the spectral dimension, e.g. for the phenomenological model of relative locality \cite{AmelinoCamelia:2011bm}, which is conjectured to carry properties of the full quantum gravity regime. In this context, it would also be of interest to investigate whether additional data carried by the causal set, which can, e.g. correspond to momentum-information as in \cite{Cortes:2013pba}, can alter the spectral dimension.
\newline

Moreover, we introduce a new probe of quantum spacetimes by considering only random walks that obey causality. The allowed moves of the random walk have to be forward in time and can only lie within the local lightcone. This new probe of quantum geometry is adapted to a Lorentzian setting, and well-suited to study causal sets. A notion of dimensionality, which we call the causal spectral dimension, can then be derived from the meeting probability of two random walkers. We discuss the causal spectral dimension for continuum flat spacetime, where it equals the topological dimension. We then show by explicit simulations, that the same is true for causal sets which are approximated by manifolds, and where the large-scale causal spectral dimension equals the topological one. The small-scale value shows an increase of the causal spectral dimension, which again is connected to the causal set nonlocality. Finally, we show an explicit example where the causal spectral dimension and the spectral dimension differ, in particular $d_{c\, s} <d_s$. In this case, this difference can be connected to the existence of a preferred time foliation in these causets. The preferred foliation implies that in time direction, the diffusion process does not lead to a spread of the probability density, which instead is concentrated on one spatial hypersurface at each step. This implies a causal spectral dimension which is smaller than the spectral dimension. 

We therefore propose the causal spectral dimension as a new, independent probe of quantum spacetime, that could be insightful to study in other Lorentzian approaches to quantum gravity. As its value can differ from the standard spectral dimension, a combination of both probes can yield enlightening insights into quantum gravity models. 
Our investigation of the causal spectral dimension is a first implementation of diffusion processes as probes of quantum spacetimes respecting causality. As such, it can be argued to be closer to physical particle propagation on quantum spacetimes. It is thus a new interesting tool to study the fundamental structure of spacetime in different quantum gravity approaches.

{\emph{Acknowledgments:}
We thank Fay Dowker, Joe Henson, Rafael Sorkin and Sumati Surya for helpful discussions. {We would like to express our thanks to Sumati Surya for a careful reading of this manuscript and for providing helpful comments.
S.~M. thanks Perimeter Institute for hospitality during this work and Girton College, Cambridge for support.
This research was supported in part by Perimeter Institute for Theoretical Physics. Research at Perimeter Institute is supported by the Government of Canada through Industry Canada and by the Province of Ontario through the Ministry of Research and Innovation.}

\end{document}